\documentclass[16pt]{article}
\usepackage{amscd}
\usepackage{bbm}
\usepackage{mathrsfs}
\usepackage{amssymb}
\usepackage{graphics}
\usepackage{graphicx}
\usepackage{amsfonts}
\usepackage{amsmath}
\usepackage{array}
\usepackage{booktabs}
\usepackage{units}

\usepackage[numbers,sort&compress]{natbib}
\begin{document}
\date{}
\title{Construction and analysis of multi-lump solutions of dispersive long wave equations via integer partitions}
\author{Yong-Ning An,  Rui Guo$\thanks{Corresponding author:
guorui@tyut.edu.cn}$\
\\
\\{\em
School of Mathematics, Taiyuan University  of} \\
{\em Technology, Taiyuan 030024, China}
} \maketitle

\begin{abstract}

In this paper, the relation between the integer partition theory and a kind of rational solution of the dispersion long wave equations is studied. For the integer partition $\lambda=\left(\lambda_1,\lambda_2,\cdots,\lambda_n\right)$ of positive integer $N$, with the degree vector $m=\left(m_1, m_2, \cdots, m_n\right)$, the corresponding $M$ lump solution can be obtained where $M=N+nm_n$. Combined with the generalized Schur polynomial and heat polynomial, the asymptotic positions of peaks are studied, and the arrangement of multi-peak groups in multi-lump solutions are obtained, as well as the relationship between the patterns formed by single-peak groups and the corresponding integer partition.

\vspace{5mm}\noindent\emph{Keywords}: The dispersive long wave equation;  Generalized Schur polynomial; Multi-lump solutions; Integer partition
\end{abstract}

\vspace{7mm}\noindent\textbf{1  Introduction}
\hspace*{\parindent}
\renewcommand{\theequation}{1.\arabic{equation}}\\

The $\left(2+1\right)$ dimensional dispersion long wave equations(DLWEs)
\begin{equation}\label{1}
\begin{aligned}
u_{yt}+\left(v_x+uu_y\right)_x&=0,\\
v_t+\left(uv-u_{xy}\right)_x&=0,
\end{aligned}
\end{equation}
where $u$ and $v$ are wave amplitude functions in relation to spatial variables ($x$ and $y$) and time variable ($t$). Eq.~(\ref{1}) is a mathematical model that describes the scenario of wide channels or open oceans with finite depth, which has substantial research significance in physics, engineering and earth science~\cite{ck81,ck82,ck83,ck84, ck85,ck86,ck87,ck1}. The DLWEs were initially introduced by Boiti in $1987$ as a compatibility condition for a weak Lax pair~\cite{ck1}. Over the past few years, various intriguing properties of the $\left(2+1\right)$-dimensional DLWEs have been extensively studied~\cite{ck31,ck32,ck33,ck34,ck35,ck36,ck37,ck38,ck39,ck40}.

In recent years, there has been a significant amount of interest in lump solutions as a new area of research. A lump is a specific object or structure characterized as a nonlinear wave, maintaining its shape and amplitude as it propagates through a medium~\cite{ck20,ck21}. The study of lumps has been extensively explored in nonlinear dynamics and has demonstrated potential applications in various areas, such as data transmission, energy storage, and quantum computing~\cite{ck22,ck23,ck24,ck25,ck26}. Furthermore, it plays a crucial role in gaining a deeper understanding of complex wave phenomena and nonlinear dynamic behaviors. Take the Kadomtsev-Petviashvili(KP) equation for instance. In Ref.~\cite{ck10}, twelve classes of lump-kink solutions to the KP equation were obtained using Hirota bilinear methods. Ref.~\cite{ck11} utilized a simplified form of a function to generate a class of generalized solutions for the KP equation and discussed different distribution characteristics of lump chains.
Ref.~\cite{ck12,ck13} used the Binary Darboux transform (BDT) and polynomial theory to develop multi-lump solutions to the KPI problem. Then they evaluated the solution by introducing the partition notion, offering significant insights into the phenomena.

This paper aims to study the lump solutions under high-order BDT. By doing so, we can enhance our comprehension of the structure and characteristics of lump solutions in DLWEs, particularly in relation to various orders and integer partitions. We can find that the different orders $n$ of the BDT and the different integer partition $\lambda=\left(\lambda_1,\lambda_2,\cdots, \lambda_n\right)$ of $N$ will correspond to the solutions with varying amounts of lump, but they can be regarded as composed of the multi-peak groups and the single-peak groups. Each single-peak group contains only one lump, while each multi-peak group contains $n$ lumps. In addition, in terms of the location of the distribution, the multi-peak groups are usually located near the coordinate axis and approximately parallel distribution, situated on the $r$-axis when $t>0$, and located near the $s$-axis when $t<0$. The distribution of single-peak groups is related to the Young diagram of integer partition of $N$. In particular, we discuss the distribution properties of single-peak groups in special partitions such as rectangular partitions, trapezoidal partitions, triangular partitions, odd partitions and even partitions with concrete examples.

The structure of this paper is as follows: In Sect.~$2$, we will introduce the generalized Schur polynomial and its simple properties, and then convert the expression of DLWEs' lump solutions constructed by BDT into the sum of squares of the modules. Lastly, we will give some examples. In Sect.~$3$, we will introduce the integer partition theory and establish its relationship with multi-lump solutions. In order to explore the peak distribution of multiple lump solutions, we will establish the relationship between generalized Schur polynomials and the heat polynomials in Sect.~$4$, especially the graphs formed by the distribution of single-peak groups in some special partitions. Finally, the conclusions will be summarized in Sect.~$5$.

\vspace{5mm} \noindent\textbf {\textbf{2  Construction of multi-lump solutions }}
\hspace*{\parindent}
\renewcommand{\theequation}{2.\arabic{equation}}\setcounter{equation}{0}
\\

In this section, we will explain the process of constructing a family of multi-lump solutions using Gramians, which are derived from the BDT applied to the DLWEs. The solutions are constructed using a specific kind of complex polynomials known as the generalized Schur polynomials, which are explained in detail below.

\vspace{5mm}
\textbf{2.1 Generalized Schur polynomials and new notation}
\\

The generic expression for the generalised Schur polynomials is defined by the relation~\cite{ck61,ck62,ck63}
\begin{equation}\label{2.1}
\begin{aligned}
\frac{1}{n!}\frac{\partial ^n\phi}{\partial k^n}e^{i\sigma}=p_n\left(k\right)e^{i\sigma},\ \ \
\sigma\left(x,y,t,k\right)=:kx+k^3y+k^2t+\gamma\left(k\right),
\end{aligned}
\end{equation}
where $k$ is a complex parameter and $\gamma\left(k\right)$ is an arbitrarily differentiable function of $k$.
By definition, we can know that $p_n$ is a polynomial in $n$ variables $\sigma_1, \sigma_2,\cdots, \sigma_n $ with $\sigma_j:=i\frac{\partial_k^j \sigma}{j!}$, which gives us
\begin{equation}\label{2.2}
\begin{aligned}
\sigma_1=i\left(x+3k^2y+2kt+\gamma_1\right),\ \ \sigma_2=i\left(3ky+t+\gamma_2\right),\ \ \sigma_3=i\left(y+\gamma_3\right),\ \ \sigma_j\left(k\right)=i\gamma_j\left(k\right),\ j>3,
\end{aligned}
\end{equation}
where $\gamma_j\left(k\right)=\frac{\partial_k^j\gamma\left(k\right)}{j!}$ are independent complex parameters related to $k$ only.

The expression of the generalised Schur polynomial $p_n$ can be obtained by comparing Eqs.~(\ref{2.3}) and~(\ref{2.4}), which are Taylor expansions of $e^{i\sigma\left(k+h\right)}$ in different orders
\begin{equation}\label{2.3}
\begin{aligned}
e^{i\sigma\left(k+h\right)}=\sum_{n=0}^{\infty}{\frac{h^n}{n!}}\frac{\partial ^n}
{\partial k^n}e^{i\vartheta \left( k \right)}=e^{i\sigma}\sum_{n=0}^{\infty}{h^np_n\left( k \right)},
\end{aligned}
\end{equation}
\begin{equation}\label{2.4}
\begin{aligned}
e^{i\sigma \left( k+h \right)}=e^{i\left( \sigma \left( k \right) +\sigma^{'}\left( k \right) h+\sigma^{''}\left( k \right) \nicefrac{h^2}{2}+\cdots \right)}=e^{i\sigma}\prod_{j=1}^{\infty}{e^{h^j\sigma _j}}=e^{i\sigma}\prod_{j=1}^{\infty}{\left[ \sum_{m_j=0}^{\infty}{\frac{\left( h^j\sigma _j \right) ^{m_j}}{m_j!}} \right]}.
\end{aligned}
\end{equation}
By comparing the two equations above in which the powers of $h$ are equal, we obtain
\begin{equation}\label{2.5}
\begin{aligned}
p_n\left(k\right)=\sum_{\begin{array}{c}
	m_1+2m_2+\cdots +nm_n=n\\
\end{array}}
{\frac{\sigma _1^{m_1}\sigma _2^{m_2}\cdots \sigma _n^{m_n}}{m_1!m_2!\cdots m_n!}},
\end{aligned}
\end{equation}
where $m_1, m_2, \cdots, m_n \ge 0$.
Based on the above equation, we get the first few terms of $p_n$ as
\begin{equation}\label{2.6}
\begin{aligned}
p_1=\sigma_1,\ \ \ p_2=\frac{\sigma_1^2}{2}+\sigma_2,\ \ \ p_3=\frac{\sigma_1^3}{3!}+\sigma_1\sigma_2+\sigma_3,\ \ \ \cdots.
\end{aligned}
\end{equation}
From Eq.~(\ref{2.5}), we know that $p_n$ is a weighted polynomial of order $n$ in $\sigma_j, j=1,\cdots,n$, where weights $wt(\sigma_j)=j$ and each $p_n$ has degree $n$ in $x, y, t$ since $\sigma_1^n$ appears in the expansion.
There is a useful property that can be derived from Eqs.~(\ref{2.1}) and~(\ref{2.2}) for the generalized Schur polynomials as follows
\begin{equation}\label{2.7}
\begin{aligned}
\frac{\partial ^jp_n}{\partial \sigma _1^j}=\frac{\partial p_n}{\partial \sigma _j}=\left\{ \begin{array}{l}
	p_{n-j},\ \ j\le n\\
	0,\ \ \ \ \ \ \ j>n\\
\end{array} \right. .
\end{aligned}
\end{equation}

Next, we present the new notations and their basic features. Let
\begin{equation}\label{2.8}
\begin{aligned}
Q_{m,n}=\left(p_{m-1}+k_0p_m\right)p_n^\ast,\ \ \ \ Q_{m,n}^\ast=\left(p_{m-1}^\ast+k_0^\ast p_m^\ast\right)p_n,
\end{aligned}
\end{equation}
where the first term of the subscript of $Q_{m,n}$ represents the maximum index of $p_n$ inside the parentheses, and the second term of the subscript represents the index of $p_n^\ast$ outside the parentheses, while the subscript in $Q_{m,n}^\ast$ has the opposite meaning.

Review before, $p_n$ is a generalized Schur polynomial with respect to $\sigma_1,\cdots,\sigma_n$, and $p_n^\ast$ is a polynomial with respect to $\sigma_1^\ast,\cdots,\sigma_n^\ast$, then we can get some simple properties of the derivative of $Q_{m,n}$ with respect to $\sigma_i, \sigma_j^\ast, \left(i=1, \cdots, m, j=1, \cdots, n \right)$.

{\bf Property 2.1} Properties of the polynomial $Q_{m,n}$.

\vspace{1mm}
\textbf{(a)} Derivative of $Q_{m,n}$ with respect to $\sigma_i\left(i=1, \cdots, m\right)$ and $\sigma_j^\ast\left(j=1, \cdots, n \right)$.
\begin{subequations}
\begin{align}
&\frac{\partial ^iQ_{m,n}}{\partial \sigma _1^i}=\frac{\partial Q_{m,n}}{\partial \sigma _i}=\left\{ \begin{array}{l}
	Q_{m-i,n},\ \ i\le m\\
	0,\ \ \ \ \ \ \ \ \ \ \ i>m\\
\end{array} \right. ,\label{2.9a}\\
&\frac{\partial ^jQ_{m,n}}{\partial \sigma _1^{\ast^j}}=\frac{\partial Q_{m,n}}{\partial \sigma _j^\ast}=\left\{ \begin{array}{l}
	Q_{m,n-j},\ \ j\le n\\
	0,\ \ \ \ \ \ \ \ \ \ \ j>n\\
\end{array} \right. ,\label{2.9b}\\	
&\frac{\partial ^{i+j}Q_{m,n}}{\partial \sigma _1^i\partial\sigma _1^{\ast^j}}=\frac{\partial^2 Q_{m,n}}{\partial \sigma _i\partial\sigma _j^\ast}=\left\{ \begin{array}{l}
	Q_{m-i,n-j},\ \ i\le m\ \text{and}\ j\le n\\
	0,\ \ \ \ \ \ \ \ \ \ \ \ \ \ i>m\ \text{or}\ j>n\\
\end{array} \right. .\label{2.9c}	
\end{align}	
\end{subequations}

\vspace{1mm}
\textbf{(b)} Derivative of $Q_{m,n}$ with respect to $x$.
\begin{equation}\label{2.10}
\begin{aligned}
\frac{\partial ^lQ_{m,n}}{\partial x^l}&=:Q_{m,n,l} =i^lZ_{m,n,l} =i^l\sum_{j=0}^{l}\tbinom{l}{j}\left(-1\right)^jQ_{m-l+j,n-j},
\end{aligned}
\end{equation}
where the third term of the subscript of $Q_{m,n,l}$ denotes the degree to which the polynomial $Q_{m,n}$ takes the derivative with respect to $x$.

\vspace{5mm}
\textbf{2.2 Expression for multi-lump solutions}
\\

Based on previous studies, we have constructed the expression $v\left(x,y,t\right)$ for the multi-lump solution of DLWEs as follows
\begin{equation}\label{2.11}
\begin{aligned}
v\left(x,y,t\right)=-2\frac{\partial^2}{\partial x\partial y}\ln \left(\left|\bf{H}\right| \right),\ \ \ \ \left|\bf{H}\right|\left(x,y,t\right)=
\left|\bf{A}\right|\left|\bf{B}\right|,
\end{aligned}
\end{equation}
where
\begin{equation}\label{2.12}
\begin{aligned}
&A_{ij}=\sum_{l=0}^{m_i+m_j}{\frac{\partial _{x}^{l}Q_{m_i,m_j}}{\left( 2b \right) ^l}},\ \ \ \ &&B_{ij}=\sum_{l=0}^{m_i+m_j}{\frac{\partial _{x}^{l}Q_{m_j,m_i}^\ast}{\left( 2b \right) ^l}},\\ &H_{ij}=\sum_{k=1}^n A_{ik}B_{kj},
&&i, j=1, 2, \cdots, n,
\end{aligned}
\end{equation}
where $n$ represents the order of the BDT. From the related content of polynomials introduced in the previous subsection, we can know that it is the term of $k=n$ that plays the main role in $H_{ij}$, and thus we obtain
\begin{equation}\label{2.13}
\begin{aligned}
H_{ij}\approx A_{in}B_{nj}
&=\sum_{\alpha=0}^{m_i+m_n}\frac{\partial_x^\alpha Q_{m_i,m_n}}{\left(2b\right)^\alpha} \sum_{\beta=0}^{m_n+m_j}\frac{\partial_x^\beta Q_{m_j,m_n}^\ast}{\left(2b\right)^\beta}\\
&=\sum_{\alpha=0}^{m_i+2m_n+m_j}\frac{1}{\left(2b\right)^\alpha} \sum_{\beta=0}^{\alpha}\partial_x^{\beta}Q_{m_i,m_n}\partial_x^{\alpha-\beta}Q_{m_j,m_n}^\ast\\
&=\sum_{\alpha=0}^{m_i+m_n}\sum_{\beta=0}^{m_n+m_j}\frac{1}{\left(2b\right)^{\alpha+\beta}}
\tbinom{\alpha+\beta}{\beta} \partial_x^{\alpha}Q_{m_i,m_n}\partial_x^{\beta}Q_{m_j,m_n}^\ast.
\end{aligned}
\end{equation}
Notice that the upper limits of both sums in the last equality above can be extended to $2m_n$ without any loss of generality, because $\partial_x^{r}Q_{m_i,m_n}=0$ if $r >m_i+m_n$. Next consider $n$ complex vectors in $\mathbb{C}^{2m_n+1}$
\begin{equation}\label{2.14}
\begin{aligned}
\textbf{P}_{i}:=\left(Q_{m_i,m_n},\partial_xQ_{m_i,m_n},\cdots, \partial_x^{2m_n}Q_{m_i,m_n}\right)^T,\ \ i=1, 2, \cdots, n,
\end{aligned}
\end{equation}
the elements of the  $(2m_n+1) \times n$ matrix can be represented as
\begin{equation}\label{2.15}
\begin{aligned}
&P_{rj}=\partial _x^rQ_{m_j,m_n}=i^rW_{rj},\\
&W_{rj}=\left\{ \begin{array}{l}
	Z_{m_j,m_n,r},\ \ r\le m_j+m_n\\
	0,\ \ \ \ \ \ \ \ \ \ \ \ r>m_j+m_n
\end{array} \right.,\ \ r=0, 1, \cdots, 2m_n,\ \ j=1, 2, \cdots, n.
\end{aligned}
\end{equation}
Then the elements of $H_{ij}$ are given by the inner products
\begin{equation}\label{2.16}
\begin{aligned}
H_{ij}=\textbf{P}_{j}^{\dag}\textbf{C}\textbf{P}_{i},\ \
C_{\alpha,\beta}=\frac{1}{\left(2b\right)^{\alpha+\beta}}\tbinom{\alpha+\beta}{\beta}, \ \ i=0, 1, 2, \cdots, 2m_n,
\end{aligned}
\end{equation}
where $\dag$ stands for conjugate transpose and $\textbf{C}$ is a real, symmetric $(2m_n+1)\times (2m_n+1)$ matrix that has a unique decomposition
\begin{equation}\label{2.17}
\begin{aligned}
\textbf{C}=\textbf{U}^{\dag}\textbf{D}\textbf{U},\ \ \ U_{rs}=\left\{ \begin{array}{l}
	\frac{1}{\left(2b\right)^{s-r}}\tbinom{s}{r},\ \ r\le s\\
	0,\ \ \ \ \ \ \ \ \ \ \ \ \ r>s\\
\end{array} \right.,\ \ r, s=0, 1, \cdots, 2m_n.
\end{aligned}
\end{equation}
Here $\textbf{U}$ is a $\left(2m_n+1\right)\times\left(2m_n+1\right)$ upper-triangular matrix with $1^{\prime}$s along its main diagonal and $\textbf{D}$ is a diagonal matrix with $D_{rr}=\left(2b\right)^{-2r},\ r=0, 1, \cdots, 2m_n$.
Then the entries of martix $\textbf{H}$ from Eq.~({\ref{2.16}) are expressed as
\begin{equation}\label{2.18}
\begin{aligned}
H_{ij}=\textbf{P}_{j}^{\dag}\textbf{C}\textbf{P}_{i}= \textbf{Q}_{j}^{\dag}\textbf{D}\textbf{Q}_{i},\ \ \textbf{Q}_{i}=\textbf{U}\textbf{P}_{i}.
\end{aligned}
\end{equation}

Let $\textbf{Q}=\textbf{UP}$, where $\textbf{P}$ is defined in Eq.~(\ref{2.15}), then the determinant of $\textbf{H}=\textbf{Q}^{\dag}\textbf{D}\textbf{Q}$ can be expressed as a sum of squares using the Cauchy-Binet formula
\begin{equation}\label{2.19}
\begin{aligned}
\tau=\text{det}\ \textbf{H}=
\sum_{\begin{array}{c}
	0\le l_1<\cdots<l_n\le m_n
\end{array}}
{\frac{\left|Q\left(l_1\cdots l_n\right)\right|^2} {\left(2b\right)^{2\left(l_1+\cdots+l_n\right)}}},\\
Q\left(l_1\cdots l_n\right)=\sum_{0\le r_1<\cdots<r_n\le m_n}
U\tbinom{l_1\cdots l_n}{r_1\cdots r_n}
P\left( r_1\cdots r_n \right),
\end{aligned}
\end{equation}
where $U\tbinom{l_1\cdots l_n}{r_1\cdots r_n}$ is the $n \times n$ minor of $\textbf{U}$ obtained from the submatrix whose rows and columns are indexed by $\left(l_1,\cdots,l_n\right)$ and $\left(r_1,\cdots,r_n\right)$, respectively. Since $\textbf{U}$ is upper-triangular matrix with $1^{\prime}$s along its main diagonal, it follows that the principal minors $U\tbinom{l_1\cdots l_n}{l_1\cdots l_n}=1$ and
$U\tbinom{l_1\cdots l_n}{r_1\cdots r_n}\ne 0$ if and only if $l_j \le r_j$ for each $j = 1,\cdots,n$.

To facilitate further discussion, we introduce the total lexicographic order for the multi-index sets $\textbf{l}:=\left(l_1\cdots l_n\right), 0\le l_1<\cdots<l_n\le m_n$.

{\bf Definition 2.1\ } Given two multi-index sets $\textbf{l}, \textbf{r}$, if $j$ is the earliest index that causes $\textbf{l}$ and $\textbf{r}$ to differ, thus $\textbf{l}< \textbf{r}$ if and only if $l_j<r_j$. We denote the smallest element $\left(01\cdots n\right)$ in the lexicographic order as $\textbf{0}$.

After using the notation above, we can express Eq.~(\ref{2.19}) in the following simplified form
\begin{equation}\label{2.20}
\begin{aligned}
\tau=\text{det}\ \textbf{H}=
\sum_{\begin{array}{c}
	\textbf{r}\ge\textbf{0}\\
\end{array}}
{\frac{\left|Q\left(\textbf{r}\right)\right|^2} {\left(2b\right)^{2\left|\textbf{r}\right|}}},\ \ \ \ \ \
Q\left(\textbf{r}\right)=\sum_{\textbf{s}\ge\textbf{r}}
U\tbinom{\textbf{r}}{\textbf{s}} P\left(\textbf{s}\right),
\end{aligned}
\end{equation}
with $\left|\textbf{r}\right|:=r_1+\cdots+r_n$. The leading polynomial term in det $\textbf{H}$ is given by ${\frac{\left|Q\left(\textbf{0}\right)\right|^2} {\left(2b\right)^{n\left(n-1\right)}}}$, where
\begin{equation}\label{2.21}
\begin{aligned}
Q\left(\textbf{0}\right)=P\left(\textbf{0}\right)+\sum_{\textbf{s}\ge\textbf{0}}
U\tbinom{\textbf{0}}{\textbf{s}}P\left(\textbf{s}\right).
\end{aligned}
\end{equation}
For any set of multiple indices $\textbf{r}$, we can get that $P\left(\textbf{r}\right)= i^{\left|\textbf{r}\right|}W\left(\textbf{r}\right)$, in particular $P\left(\textbf{0}\right)= i^{\frac{n\left(n-1\right)}{2}}W\left(\textbf{0}\right)$, where
\begin{equation}\label{2.22}
\begin{aligned}
W\left( \textbf{r} \right) =\left| \begin{matrix}
	Z_{m_1,m_n,r_1}&		Z_{m_2,m_n,r_1}&		\cdots&		Z_{m_n,m_n,r_1}\\
	Z_{m_1,m_n,r_2}&		Z_{m_2,m_n,r_2}&		\cdots&		Z_{m_n,m_n,r_2}\\
	\vdots&		\vdots&		\vdots&		\vdots\\
	Z_{m_1,m_n,r_n}&		Z_{m_2,m_n,r_n}&		\cdots&		Z_{m_n,m_n,r_n}\\
\end{matrix} \right|,\ \
W\left( \textbf{0} \right) =\left| \begin{matrix}
	Z_{m_1,m_n}&		Z_{m_2,m_n}&		\cdots&		Z_{m_n,m_n}\\
	Z_{m_1,m_n,1}&		Z_{m_2,m_n,1}&		\cdots&		Z_{m_n,m_n,1}\\
	\vdots&		\vdots&		\vdots&		\vdots\\
	Z_{m_1,m_n,n-1}&		Z_{m_2,m_n,n-1}&		\cdots&		Z_{m_n,m_n,n-1}\\
\end{matrix} \right|
\end{aligned}
\end{equation}
are the $n\times n$ minors of the $\left(2m_n+1\right)\times n$ matrix $\textbf{W}$ defined in Eq.~({\ref{2.15}). Notice that $P\left(\textbf{r}\right)$ is a weighted homogeneous polynomial in $\sigma_1, \cdots, \sigma_{m_n}, \sigma_1^\ast, \cdots, \sigma_{m_n}^\ast$ of degree $\left(m_1+\cdots +m_n\right)+nm_n-\left|\textbf{r}\right|$, and which of $P\left(\textbf{0}\right)$ is $M:=\left(m_1+\cdots +m_n\right)+nm_n-\frac{n\left(n-1\right)}{2}$.

\vspace{5mm}
\textbf{2.3 Examples of multi-lump solutions}
\\

In this subsection, we will give several examples of multi-lump solutions of the DLWEs. For simplicity, we will introduce the following set of coordinates
\begin{equation}\label{2.23}
\begin{aligned}
r=x^{\prime}+3\left(a^2-b^2\right)y^{\prime},\ s=6aby^{\prime}, \text{where}\ \ x^{\prime}=x+\frac{a^2+b^2}{a}t,\ y^{\prime}=y+\frac{1}{3a}t.
\end{aligned}
\end{equation}
Recall that $a=Re\left(k_0\right)$ and $b=Im\left(k_0\right)$, let $a=\omega b$ and parameter $b$ be a constant, so that $k_0=b\left(\omega+i\right)$ can be obtained. By changing the value of $\omega$, we can know the effects of $k_0$ on DLWEs' lump solutions. Thus, the variable $\sigma_j$ in Eq.~(\ref{2.2}) becomes the following
\begin{equation}\label{2.24}
\begin{aligned}
\sigma_1=i\left(r+is+\gamma_1\right),\ \ \sigma_2=-\frac{s}{2b\omega}+\frac{t}{\omega}+\frac{is}{2b}+i\gamma_2,\ \ \sigma_3=\frac{is}{6b^2\omega}-\frac{it}{3b\omega}+i\gamma_3,\ \ \sigma_j\left(k\right)=0,\ j>3,
\end{aligned}
\end{equation}
where we can set $\gamma_j=0$ for simplicity.

\vspace{5mm}
\textbf{Example 2.1} Let's first consider the case $n=2$, where $m_1=1$ and $m_2=2$. In this case, the matrices $\textbf{P}, \textbf{U}$ and $\textbf{D}$ are given by
\begin{equation}\label{2.25}
\begin{aligned}
\textbf{P}=\left( \begin{array}{cc}
	Z_{1,2}&		Z_{2,2}\\
	Z_{1,2,1}&		Z_{2,2,1}\\
	Z_{1,2,2}&		Z_{2,2,2}\\
	Z_{1,2,3}&		Z_{2,2,3}\\
	Z_{1,2,4}&		Z_{2,2,4}\\
\end{array} \right),\ \ \
\textbf{U}=\left( \begin{array}{ccccc}
	1&		\frac{1}{2b}&		\frac{1}{4b^2}&		\frac{1}{8b^3}&		\frac{1}{16b^4}\\
	0&		1&		\frac{1}{b}&		\frac{3}{4b^2}&		\frac{1}{2b^3}\\
	0&		0&		1&		\frac{3}{2b}&		\frac{3}{2b^2}\\
	0&		0&		0&		1&		\frac{2}{b}\\
	0&		0&		0&		0&		1\\
\end{array} \right),\ \ \ \textbf{D}=\text{diag}\left(1,\frac{1}{(2b)^2}, \frac{1}{(2b)^4},\frac{1}{(2b)^6},\frac{1}{(2b)^8}\right).
\end{aligned}
\end{equation}
Next, calculating all the $2\times 2$ minors using Eq.~(\ref{2.20}) leads to the sum of squares expression
\begin{equation}\label{2.26}
\begin{aligned}
\tau&=\frac{\left|Q\left(01\right)\right|^2}{\left(2b\right)^2}+
\frac{\left|Q\left(02\right)\right|^2}{\left(2b\right)^4}+
\frac{\left|Q\left(03\right)\right|^2+\left|Q\left(12\right)\right|^2}{\left(2b\right)^6}+
\frac{\left|Q\left(04\right)\right|^2+\left|Q\left(13\right)\right|^2}{\left(2b\right)^8}\\
&+\frac{\left|Q\left(14\right)\right|^2+\left|Q\left(23\right)\right|^2}{\left(2b\right)^{10}}+
\frac{\left|Q\left(24\right)\right|^2}{\left(2b\right)^{12}}+
\frac{\left|Q\left(34\right)\right|^2}{\left(2b\right)^{14}}.
\end{aligned}
\end{equation}
By substituting the above formula into Eq.~(\ref{2.11}), we get the expression of the multi-lump solution of DLWEs.

\begin{center}
\includegraphics[scale=0.4]{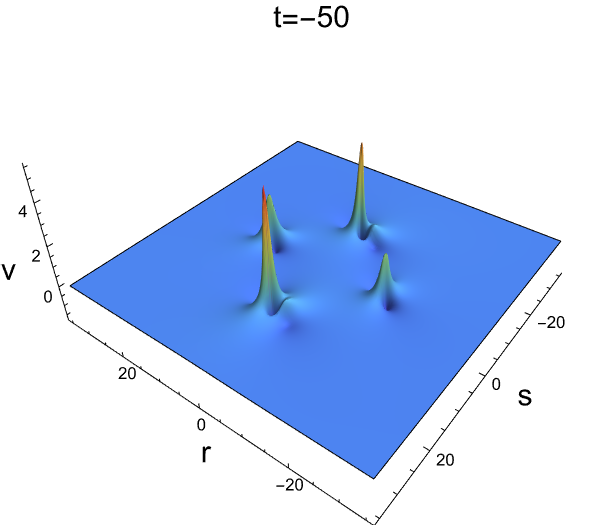}\hfill
\includegraphics[scale=0.4]{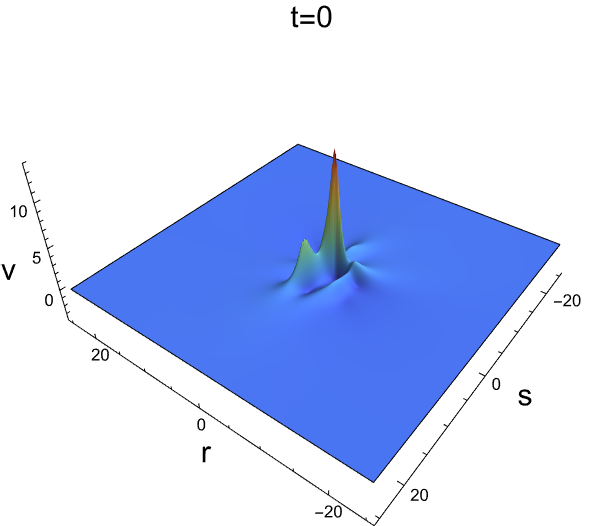}\hfill
\includegraphics[scale=0.4]{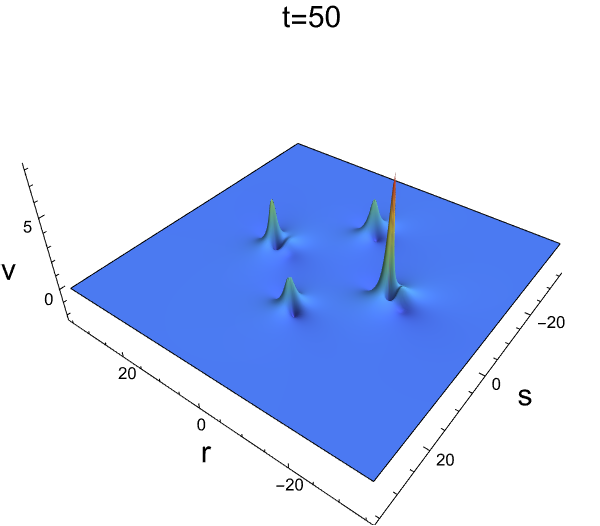}\hfill
\vspace{-0.1cm}{\footnotesize\hspace{2.8cm}(a)\hspace{5.1cm}(b)\hspace{5.1cm}(c)}\\\vspace{0.3cm}
\flushleft{\footnotesize
\textbf{Fig.~$1$.} (a)$\sim$(c) are the $6$-lump solutions of the DLWEs with $b=0.5, \omega=0.5.$. }
\end{center}

We can see from Fig.~$1$ that in this case, there are six peaks in the whole system, of which two groups contain two lumps, known as multi-peak groups, and the other two groups are single, known as single-peak groups. When $t<0$, the single-peak groups are approximately located on the $r$-axis, and the multi-peak groups are approximately located on the $s$-axis, and the peaks in these groups are parallel to the $r$-axis. As $t \rightarrow 0$, all peaks attract each other and move toward the origin. Then the deflection occurs, and at $t>0$, the peaks in the multi-peak groups are approximately parallel to the $r$-axis, and the peaks in the single-peak groups are approximately located on the $s$-axis.

\vspace{5mm}
\textbf{Example 2.2} Now let's keep $n=2$ the same as in the previous example, and set $\left(m_1,m_2\right)=\left(1,3\right)$ and $\left(2,3\right)$, respectively. We first consider the case $n=2$, where $m_1=1$ and $m_2=2$. By referring to Eqs.~(\ref{2.14}) and~(\ref{2.17}) and the same process as in Ex.~$2.1$, we can get the matrix $\textbf{P}, \textbf{U}$ and $\textbf{D}$, respectively, and thus get the result shown in Fig.~$2$.

\begin{center}
\includegraphics[scale=0.4]{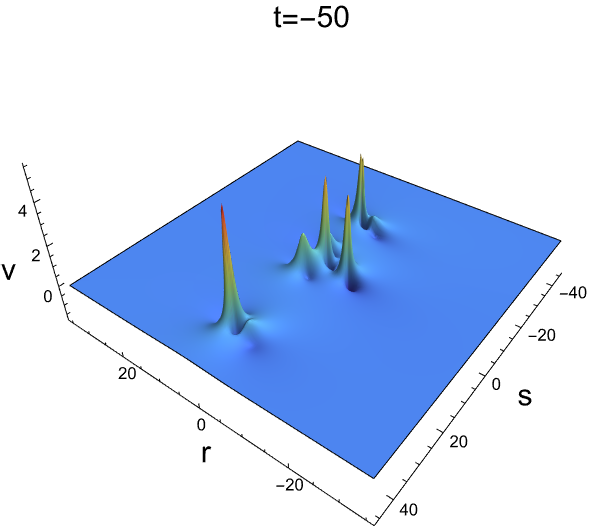}\hfill
\includegraphics[scale=0.4]{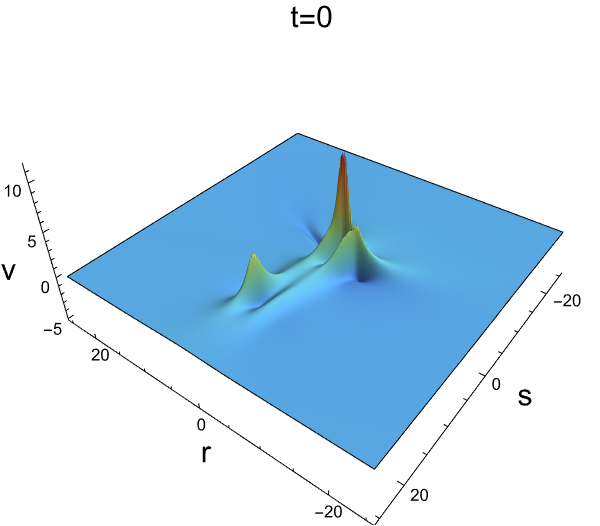}\hfill
\includegraphics[scale=0.4]{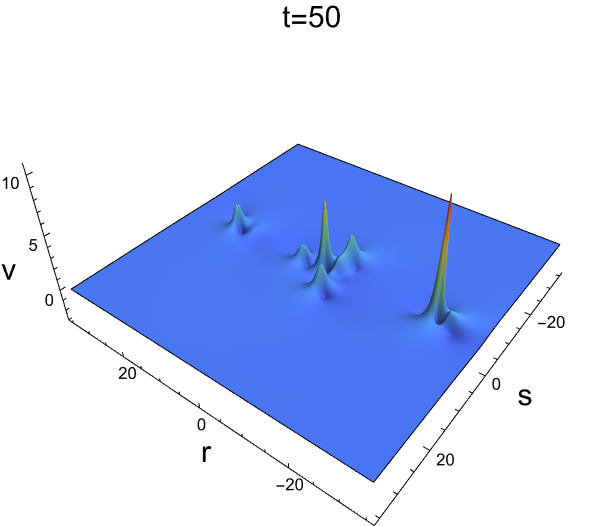}\hfill
\vspace{-0.1cm}{\footnotesize\hspace{2.8cm}(a)\hspace{5.1cm}(b)\hspace{5.1cm}(c)}\\\vspace{0.3cm}
\includegraphics[scale=0.4]{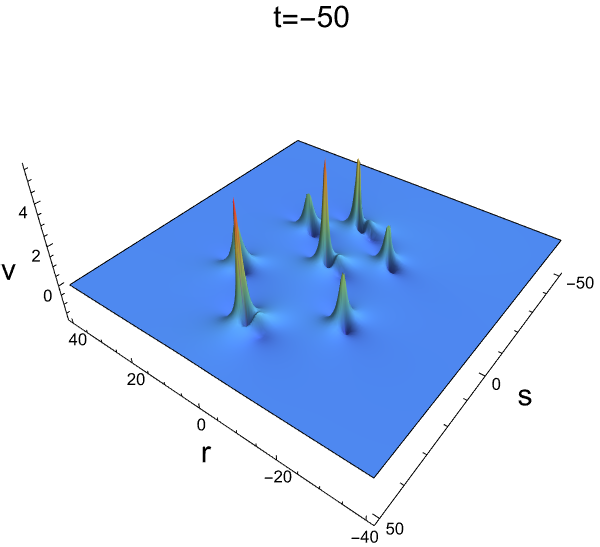}\hfill
\includegraphics[scale=0.4]{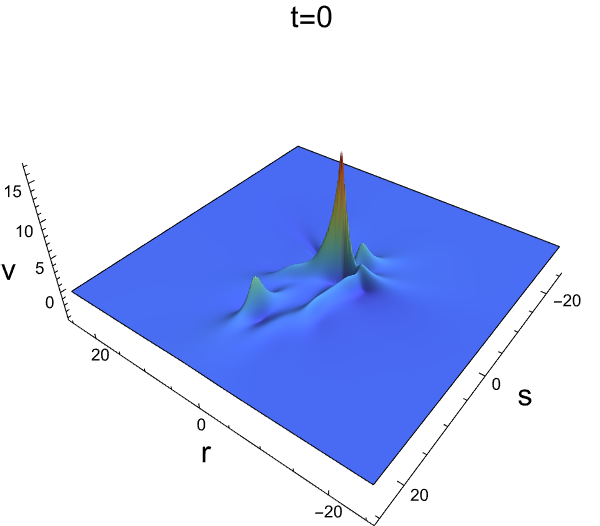}\hfill
\includegraphics[scale=0.4]{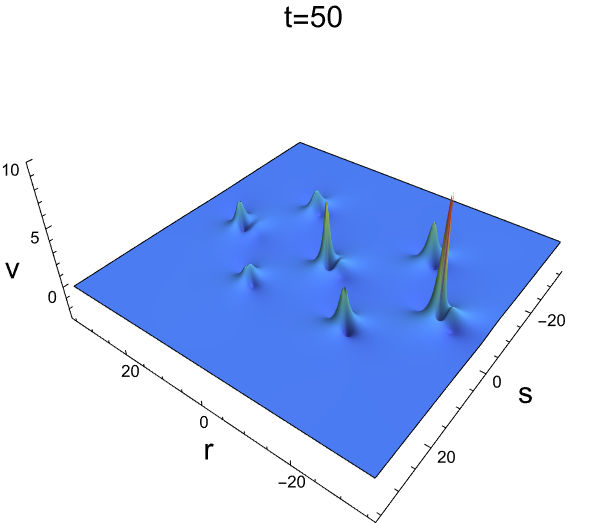}\hfill
\vspace{-0.1cm}{\footnotesize\hspace{2.8cm}(d)\hspace{5.1cm}(e)\hspace{5.1cm}(f)}\\\vspace{0.3cm}
\flushleft{\footnotesize
\textbf{Fig.~$2$.} The parameters are selected as $b=0.5, \omega=0.5.$ (a)$\sim$(c) are the $9$-lump solutions of the DLWEs. (d)$\sim$(f) are the $10$-lump solutions of the DLWEs.}
\end{center}

From Fig.~$2$(a)-(c), we can see that the whole system contains $9$ lumps, which are distributed around the $s$-axis when $t<0$, and around the $r$-axis when $t>0$. In the process of time evolution, as in the case of $6$-lump decomposition, the interiors of the system attracted to each other and then repelled away from each other. The difference, however, is that in this case, there are five lumps near the origin, and three of them can be seen as triangular distributions. That is, the entire system consists of three multi-peak groups consisting of two peaks, and three single-peak groups forming a triangular distribution.
Fig.~$2$(d)-(f) shows the $10$-lump solutions at different values of $t$, which consist of three multi-peak groups consisting of two peaks, and four single-peak groups forming a rectangular distribution. The general motion laws are consistent with the above examples.

\vspace{5mm} \noindent\textbf {\textbf{3 Integer partition theory and multi-lump solutions}}
\hspace*{\parindent}
\renewcommand{\theequation}{3.\arabic{equation}}\setcounter{equation}{0}
\\

In the previous section, we have constructed the multi-lump solutions of DLWEs, that is Eq.~(\ref{2.20}), by using the BDT and the related content of the generalized Schur polynomials. Given a set of different positive integer indices $\left\{m_1, m_2, \cdots, m_n\right\}$, we can get the corresponding expression. Therefore, in order to establish the correlation between multi-lump solutions and integer partitions, we will begin by introducing some basic ideas from partition theory and its association with polynomials.

{\bf Definition 3.1} For a given non-negative integer $N$, we decompose it and write the result as $\lambda =\left(\lambda_1, \lambda_2, \cdots, \lambda_n\right)$ in non-decreasing order such that $0\le\lambda_1\le\lambda_2\le\cdots\le\lambda_n$ and $\left|\lambda\right|:=\lambda_1+\cdots +\lambda_n=N$ satisfy. The number of non-zero parts $l\left( \lambda \right) \le n$ of $\lambda$ is called its length and $\left|\lambda\right|=N$. The strictly increasing sequence $m=\left(m_1, m_2, \cdots, m_n\right)$ defined by $m_j =\lambda_j +\left(j-1\right), j = 1,\cdots, n$ is called the degree vector of the partition $\lambda$. Similar to the above, we define the $\overline{\lambda} =\left(\lambda_1+m_n, \lambda_2+m_n, \cdots, \lambda_n+m_n\right)$,
$\overline{m}=\left(m_1+m_n, m_2+m_n, \cdots, 2m_n\right)$, then obtain that $\left|\overline{\lambda}\right|:=\lambda_1+\cdots +\lambda_n+nm_n=N+nm_n$ and $\overline{m_j} =\overline{\lambda_j} +\left(j-1\right), j = 1,\cdots, n$.

It is convenient to describe the partition $\lambda$ of the positive integer $N$ with the corresponding Young diagram $Y_\lambda$, which is a left-justified rectangular array such that the $i$th row from the top contains $\lambda_n-i+1$ boxes, $i=1, \cdots , n$.
So $Y_\lambda$ contains $n$ rows, where the $l\left(\lambda\right)$ row contains non-zero boxes, $\lambda_n$ columns, and the total number of boxes $Y_\lambda$ contains is $N$.
If we flip the Young diagram $Y_\lambda$ by interchanging its rows and columns and is called $Y_{\lambda^{\prime}}$, the partition is called the conjugate partition $\lambda^{\prime}$ of $\lambda$. Clearly, $\left|\lambda\right|=\left|\lambda^{\prime}\right|=N$, $\left(\lambda^\prime\right)^\prime=\lambda$, the length $l\left(\lambda^\prime\right)=\lambda_n$, and $\lambda^\prime$ has $l\left(\lambda\right)\le n$ columns.

For a given partition $\mu$ and $\lambda$, $\mu \subset \lambda$ means that $Y_\mu \subset Y_\lambda$, that is $\mu^i \le \lambda^i, 1\le i\le \left(\mu\right)$ where $\mu^i$ and $\lambda^i$ are the number of boxes in the $i$th row of $Y_\mu$ and $Y_\lambda$, respectively. Suppose $\lambda$ has $n$ parts, then $\mu$ can reach $n$ parts by adding a number of $0$ parts, in which case $\mu\subset\lambda$ reveals $\mu_i<\lambda_i, i=1, \cdots, n$. The set difference $Y_\lambda - Y_\mu$ is called a skew diagram $Y_{\lambda / \mu}$ which may be disconnected.

\vspace{5mm}
\textbf{Example 3.1} Let $\lambda=\left(1,1,3,4\right), \mu=\left(1,2,3\right)$, then we have
\\
\vspace{0.8cm}
$Y_\lambda=$
\begin{minipage}{0.2\textwidth}
\includegraphics[width=\textwidth]{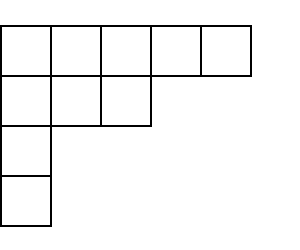}
\end{minipage}
$Y_\mu=$
\begin{minipage}{0.2\textwidth}
\includegraphics[width=\textwidth]{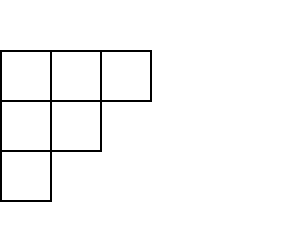}
\end{minipage}
$Y_{\lambda^\prime}=$
\begin{minipage}{0.2\textwidth}
\includegraphics[width=\textwidth]{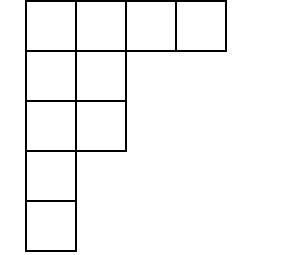}
\end{minipage}
$Y_{\lambda / \mu}=$
\begin{minipage}{0.2\textwidth}
\includegraphics[width=\textwidth]{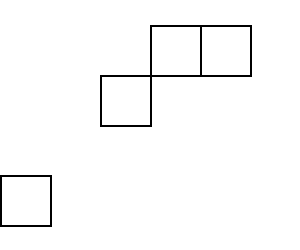}
\end{minipage}

In conjunction with Eq.~(\ref{2.22}), for each partition $\lambda$ of $N$ and the corresponding degree vector $m=\left(m_1, m_2, \cdots, m_n\right)$, we get $W\left(\textbf{0}\right)=\text{det}\left(Z_{m_j,m_n,i-1}\right)= \text{det}\left(Z_{\lambda_j+\left(j-1\right),\lambda_n+\left(n-1\right),i-1}\right)$. Similarly, the corresponding skew Schur function can be defined as follows. Let $\lambda$ and $\mu$ be two partitions satisfying $\lambda\subset\mu$ and the degree vectors are $m$ and $n$, where $m_i=\lambda_i+\left(i-1\right)$, $n_i=\mu_i+\left(i-1\right)$, $i=1, \cdots, n$, respectively, then the corresponding skew Schur function can be expressed as
\begin{equation}\label{3.1}
\begin{aligned}
&W_\lambda=W\left(\textbf{0}\right)=\text{det}\left(Z_{m_j,m_n,i-1}\right),\\
&W_{\lambda / \mu}=\text{det}\left(Z_{m_j,m_n,i-1}\right)= \text{det}\left(Z_{\lambda_j+\left(j-1\right),\lambda_n+\left(n-1\right),\mu_i+i-1}\right).
\end{aligned}
\end{equation}
Both $W_\lambda$ and $W_{\lambda/\mu}$ are weighted homogeneous polynomials in the $\sigma_j$s and $\sigma_j^\ast$s of degree $nm_n+\left|\lambda\right|= \left|\overline{\lambda}\right|$ and $ \left|\overline{\lambda}\right|-\left|\mu\right|$ respectively, then subtract $nm_n$ from the corresponding degrees, which is the number of boxes corresponding to the Young diagram $Y_\lambda$ and the skew diagram $Y_{\lambda/\mu}$ respectively.

Notice that the multi-index set $\textbf{r}=\left(r_1\cdots r_n\right)$ with $0\le r_1< r_2<\cdots <r_n\le m_n$ introduced as a degree vector of a partition denoted by $\lambda\left(\textbf{r}\right)$. Each element $\lambda_i\left(\textbf{r}\right)=r_i-i+1$ for $1\le i\le n$ and $\left|\lambda\left(\textbf{r}\right)\right|= \left|\textbf{r}\right|-\left|\textbf{0}\right|$. In particular, if $\textbf{r}=\textbf{0}$, the partition $\lambda\left(\textbf{0}\right)=\emptyset$ has only $0$ parts, and $\lambda\left(\textbf{m}\right)=\lambda$ when $\textbf{r}=\textbf{m}$ is the last multi-index in the lexicographic order. In addition, corresponding to the lexicographic order in Definition $2.1$, we can introduce a partial order relation
\begin{equation}\label{3.2}
\begin{aligned}
\textbf{r}\preccurlyeq \textbf{s}\Longleftrightarrow r_i\le s_i,\ \ \  1\le i\le n,
\end{aligned}
\end{equation}
for multi-index sets that also satisfies $\lambda(\textbf{r})\subset\lambda(\textbf{s})$ and $Y_{\lambda(\textbf{r})}\subset Y_{\lambda(\textbf{s})}$.

The main purpose of partition theory is to connect the function $\tau$ in Eq.~(\ref{2.20}) with the Schur and skew Schur functions. We have already mentioned that in Eq.~(\ref{3.1}), that is the second formula of Eq.~(\ref{2.22}). For a given multi-index set $\textbf{r}$ and the corresponding partition $\mu$, the skew Schur function can be expressed as $W_{\lambda/\mu}=W_{\lambda/\lambda\left(r\right)}=W\left(\textbf{r}\right)$, where $W\left(\textbf{r}\right)$ is the first expression in Eq.~(\ref{2.22}). By combining Eq.~(\ref{2.21}), we can get
\begin{equation}\label{3.3}
\begin{aligned}
Q\left(\textbf{0}\right)=i^{\left|\textbf{0}\right|}\big(W_\lambda+ \sum_{\lambda\left(\textbf{s}\right)\ne\emptyset}i^{\left|\lambda\left(\textbf{s}\right)\right|} U\tbinom{\textbf{0}}{\textbf{s}}W_{\lambda/\lambda\left(s\right)}\big),
\end{aligned}
\end{equation}
where the sum is all non-empty partitions $\lambda\left(\textbf{s}\right)$. Similarly, for each $\textbf{r}>\textbf{0}$, the second formula in Eq.~(\ref{2.20}) can be written as
\begin{equation}\label{3.4}
\begin{aligned}
Q\left(\textbf{r}\right)=i^{\left|\textbf{r}\right|}\big(W_{\lambda/\lambda\left(r\right)}+ \sum_{\lambda\left(\textbf{s}\right)}i^{\left|\lambda\left(\textbf{s}\right)\right|- \left|\lambda\left(\textbf{r}\right)\right|} U\tbinom{\textbf{r}}{\textbf{s}}W_{\lambda/\lambda\left(s\right)}\big),\ \ \
\lambda\left(r\right)\subset\lambda\left(s\right)\subseteq\overline{\lambda},
\end{aligned}
\end{equation}
where the sum is for all partitions of the Young diagrams that satisfy $Y_{\lambda\left(\textbf{r}\right)}\subset Y_{\lambda\left(\textbf{s}\right)}\subseteq Y_{\overline{\lambda}}$.

\textbf{Example 3.1} Take the $6$-lump solution in Ex.~$2.1$ as an example. In this case, the partition corresponding to the degree vector $m=\left(1,2\right)$ is $\lambda=\left(1,1\right)$. Then, the $2\times 2$ minors of $P, Q, U, D$ can be labeled by the multi-index set $\textbf{r}=\{\left(r_1, r_2\right), 0\le r_1<r_2\le m_n\}$. So the Schur function $W_{\left(1,2\right)}=W\left(\textbf{0}\right)$ and the skew Schur function $W_{\lambda /\lambda\left(\textbf{r}\right)}=W\left(\textbf{r}\right)$ in this example can be denoted as
\begin{equation}\label{3.5}
\begin{aligned}
W\left(\left(0,1\right)\right)=W_{\left(1,2\right)}=\left| \begin{matrix}
	Z_{1,2}&		Z_{2,2}\\
	Z_{1,2,1}&		Z_{2,2,1}\\
\end{matrix} \right|,\ \ \ \ \
W\left(\left(0,2\right)\right)=W_{\left(1,2\right)/\left(0,1\right)}=\left| \begin{matrix}
	Z_{1,2}&		Z_{2,2}\\
	Z_{1,2,2}&		Z_{2,2,2}\\
\end{matrix} \right|,\nonumber\\
W\left(\left(0,3\right)\right)=W_{\left(1,2\right)/\left(0,2\right)}=\left| \begin{matrix}
	Z_{1,2}&		Z_{2,2}\\
	Z_{1,2,3}&		Z_{2,2,3}\\
\end{matrix} \right|,\ \ \ \ \
W\left(\left(0,4\right)\right)=W_{\left(1,2\right)/\left(0,3\right)}=\left| \begin{matrix}
	Z_{1,2}&		Z_{2,2}\\
	Z_{1,2,4}&		Z_{2,2,4}\\
\end{matrix} \right|,\nonumber\\
W\left(\left(1,2\right)\right)=W_{\left(1,2\right)/\left(1,1\right)}=\left| \begin{matrix}
	Z_{1,2,1}&		Z_{2,2,1}\\
	Z_{1,2,2}&		Z_{2,2,2}\\
\end{matrix} \right|,\ \ \ \ \
W\left(\left(1,3\right)\right)=W_{\left(1,2\right)/\left(1,2\right)}=\left| \begin{matrix}
	Z_{1,2,1}&		Z_{2,2,1}\\
	Z_{1,2,3}&		Z_{2,2,3}\\
\end{matrix} \right|,\nonumber\\
W\left(\left(1,4\right)\right)=W_{\left(1,2\right)/\left(1,3\right)}=\left| \begin{matrix}
	Z_{1,2,1}&		Z_{2,2,1}\\
	Z_{1,2,4}&		Z_{2,2,4}\\
\end{matrix} \right|,\ \ \ \ \
W\left(\left(2,3\right)\right)=W_{\left(1,2\right)/\left(2,2\right)}=\left| \begin{matrix}
	Z_{1,2,2}&		Z_{2,2,2}\\
	Z_{1,2,3}&		Z_{2,2,3}\\
\end{matrix} \right|,\nonumber\\
W\left(\left(2,4\right)\right)=W_{\left(1,2\right)/\left(2,3\right)}=\left| \begin{matrix}
	Z_{1,2,2}&		Z_{2,2,2}\\
	Z_{1,2,4}&		Z_{2,2,4}\\
\end{matrix} \right|,\ \ \ \ \
W\left(\left(3,4\right)\right)=W_{\left(1,2\right)/\left(3,3\right)}=\left| \begin{matrix}
	Z_{1,2,3}&		Z_{2,2,3}\\
	Z_{1,2,4}&		Z_{2,2,4}\\
\end{matrix} \right|.\nonumber\\
\end{aligned}
\end{equation}
Therefore, in contrast to Eq.~(\ref{3.3}), the first term in Eq.~(\ref{2.26}) can be represented as
\begin{equation}\label{3.6}
\begin{aligned}
Q\left(\textbf{0}\right)=Q\left(\left(01\right)\right)&=i[ W_{\left(1,2\right)}+\frac{i}{b}W_{\left(1,2\right)/\left(0,1\right)} -\frac{3}{4b^2}W_{\left(1,2\right)/\left(0,2\right)} -\frac{i}{2b^3}W_{\left(1,2\right)/\left(0,3\right)}\ \nonumber\\
&-\frac{1}{4b^2}W_{\left(1,2\right)/\left(1,1\right)} -\frac{i}{4b^3}W_{\left(1,2\right)/\left(1,2\right)}
+\frac{3}{16b^4}W_{\left(1,2\right)/\left(1,3\right)}\ \nonumber\\ &+\frac{1}{16b^4}W_{\left(1,2\right)/\left(2,2\right)}
+\frac{i}{16b^5}W_{\left(1,2\right)/\left(2,3\right)} -\frac{1}{16b^6}W_{\left(1,2\right)/\left(3,3\right)}],\ \nonumber\\
\end{aligned}
\end{equation}
Similarly, from Eq.~(\ref{3.4}), we can represent other terms in Eq.~(\ref{2.26}), such as \begin{equation}\label{3.7}
\begin{aligned}
Q\left(\left(02\right)\right)&=-[W_{\left(1,2\right)/\left(0,1\right)} +\frac{3i}{2b}W_{\left(1,2\right)/\left(0,2\right)} -\frac{3}{2b^2}W_{\left(1,2\right)/\left(0,3\right)}
+\frac{i}{2b}W_{\left(1,2\right)/\left(1,1\right)} -\frac{3}{4b^2}W_{\left(1,2\right)/\left(1,2\right)}\ \nonumber\\
&-\frac{3i}{4b^3}W_{\left(1,2\right)/\left(1,3\right)} -\frac{i}{4b^3}W_{\left(1,2\right)/\left(2,2\right)}
+\frac{5}{16b^4}W_{\left(1,2\right)/\left(2,3\right)} +\frac{3i}{32b^5}W_{\left(1,2\right)/\left(3,3\right)}].\ \nonumber\\
\end{aligned}
\end{equation}

Let $\lambda =\left(\lambda_1, \lambda_2, \cdots, \lambda_n\right)$ be a partition of a positive integer $N$ and $W_\lambda$ be the associated Schur function. Combining the above contents with Eqs.~(\ref{2.20}),~(\ref{3.3}) and~(\ref{3.4}), we can get the following expressed as a sum of squares of the modules
\begin{equation}\label{3.8}
\begin{aligned}
\tau &=\frac{1}{\left(2b\right)^{2\left|\textbf{0}\right|}}\left|W_\lambda+ \sum_{\lambda\left(\textbf{s}\right)\ne\emptyset}i^{\left|\lambda\left(\textbf{s}\right)\right|} U\tbinom{\textbf{0}}{\textbf{s}}W_{\lambda/\lambda\left(s\right)}\right|^2\\
&+\sum_{\textbf{r}>\textbf{0}}\frac{1}{\left(2b\right)^{2\left|\textbf{r}\right|}} \left|W_{\lambda/\lambda\left(r\right)}+ \sum_{\lambda\left(r\right)\subset\lambda\left(s\right)\subseteq\overline{\lambda}} i^{\left|\lambda\left(\textbf{s}\right)\right|- \left|\lambda\left(\textbf{r}\right)\right|} U\tbinom{\textbf{r}}{\textbf{s}}W_{\lambda/\lambda\left(s\right)}\right|^2,
\end{aligned}
\end{equation}
where $\textbf{r}=\left(r_1\cdots r_n\right)$, $0\le r_1 <\cdots <r_n \le 2m_n$ is the degree vector of the partition $\lambda\left(\textbf{r}\right)$.

\vspace{5mm} \noindent\textbf {\textbf{4 Asymptotic analysis of multi-lump solutions and special integer partitions}}
\hspace*{\parindent}
\renewcommand{\theequation}{4.\arabic{equation}}\setcounter{equation}{0}
\\

In the previous content, we have seen that the multi-lump solutions of DLWEs can be identified by the partition $\lambda$ of the positive integer $N$. To avoid redundancy, we consider in this section only those partitions $\lambda$ with the smallest element $\lambda_1>0$, that is, partitions with partition length $l\left(\lambda\right)=N$.
Similar to Definition $2.1$, let $\lambda$ and $\mu$ be two different partitions of $N$, and if $j$ is the first index to make $\mu_j\ne \lambda_j$, then there is $\mu < \lambda$ if and only if $\mu_j < \lambda_j$. Note that the lexicographic ordering is a total ordering. The smallest partition $\lambda = \left(1, 1, \cdots, 1\right) :=\left(1^N\right)$ which has $N$ parts, and the largest partition $\lambda = \left(N\right)$ that has only one part.

In the following, we will gain further insight into the dynamical properties of DLWEs by studying the evolution of the peak positions of the multi-lump solutions.
From Eq.~(\ref{2.11}), we can get
\begin{equation}\label{4.1}
\begin{aligned}
v\left(x,y,t\right)= -2\frac{\tau_{xy}\tau-\tau_{x}\tau_{y}}{ \tau^2}.
\end{aligned}
\end{equation}
When $\tau$ takes the minimum value, there are $\tau_{x}=0$ and $\tau_{y}=0$, we can get $\left|\tau\right|= 2\left|\frac{\tau_{xy}}{\tau}\right|$ for the maximum value which is the peak height.
In combination with Eq.~(\ref{3.5}), we set the minimum value of its first term, that is, let
\begin{equation}\label{4.2}
\begin{aligned}
l_0=W_\lambda+ \sum_{\lambda\left(\textbf{s}\right)\ne\emptyset} i^{\left|\lambda\left(\textbf{s}\right)\right|} U\tbinom{\textbf{0}}{\textbf{s}} W_{\lambda/\lambda\left( s\right)}=0,
\end{aligned}
\end{equation}
so that the above condition is established.

Next, recall from Eq.~(\ref{2.24}), the $t$-dependence in $p_N$ occurs via $\sigma_2$ and $\sigma_3$ which are linear in $t$. Then from Eq.~(\ref{2.5}),
\begin{equation}\label{4.3}
\begin{aligned}
p_N\left(r,s,t \right) =\sum_{m_1,m_2,m_3\ge0}{\frac{\sigma _1^{m_1}\sigma _2^{m_2}\sigma_3^{m_3}} {m_1!m_2!m_3!}},\ \ \ m_1+2m_2+3m_3=N,
\end{aligned}
\end{equation}
since $\sigma_j$ for $j>3$ are constants. In order to get the highest order in the polynomial concerning $t$, we need to maximize $m_2+m_3$ in the case of $m_1+2m_2+3m_3=N$.
From Eq.~(\ref{3.4}), $\sigma_1=i z$ where $z=r+is, \sigma_2\sim i^2\left(-\frac{t}{\omega}\right)$ and $\sigma_3\sim i^3\left(\frac{t}{2b\omega}\right)$,
then we obtain
\begin{equation}\label{4.4}
\begin{aligned}
p_n\left(r,s,t \right) \sim i^{n}\left|t\right|^{\frac{n}{2}} \Big( H_n\left(\rho,\nu\right)+\left|t\right|^{-\frac{1}{2}} \frac{1}{3b\omega} H_{n-3}\left(\rho,\nu\right)+O\big(\left|t\right|^{-1}\big)\Big),
\end{aligned}
\end{equation}
where $H_n\left(\rho,\nu\right)$ is the heat polynomial which can be obtained via the generating function~\cite{ck64,ck65,ck66}
\begin{equation}\label{4.5}
\begin{aligned}
\text{exp}\left(\alpha\rho+\alpha^2\nu\right)=\sum_{r=0}^\infty H_r\left(\rho,\nu\right)\alpha^r,
\end{aligned}
\end{equation}
and can be rewritten as
\begin{equation}\label{4.6}
\begin{aligned}
H_r\left(\rho, \nu\right)=\sum_{j,k\ge0}\frac{\rho^j\nu^k}{j!k!},\ \ \ \ j+2k=r,\ \ \ \
\nu=\left\{ \begin{array}{c}
\frac{1}{\omega} ,\ \ \ \ t<0\\
-\frac{1}{\omega},\ \ t\ge0\\
\end{array} \right..
\end{aligned}
\end{equation}
If we set $l_0=0$ in Eq.~(\ref{4.2}), then the peak locations are given by
\begin{equation}\label{4.7}
\begin{aligned}
z_j\left(t\right)=r_j\left(t\right)+is_j\left(t\right)=\left|t\right|^{\frac{1}{2}}\rho_j,\ \ \ \ j=1, 2, \cdots, N+nm_n,
\end{aligned}
\end{equation}
which depend only on $\rho$.

Then, combined with the above content, we give the peak position schematic diagrams of the cases given in Ex.~$2.1$ and~$2.2$.

\begin{center}
\includegraphics[scale=0.45]{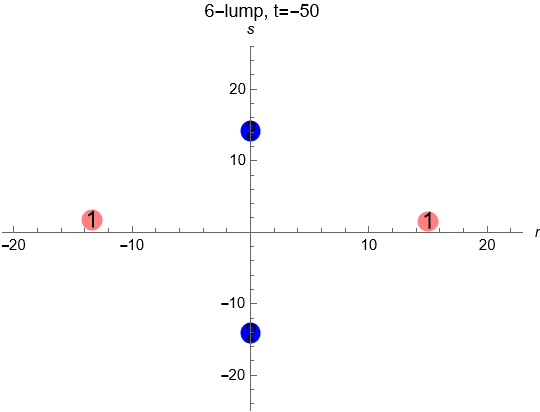}\hfill
\includegraphics[scale=0.45]{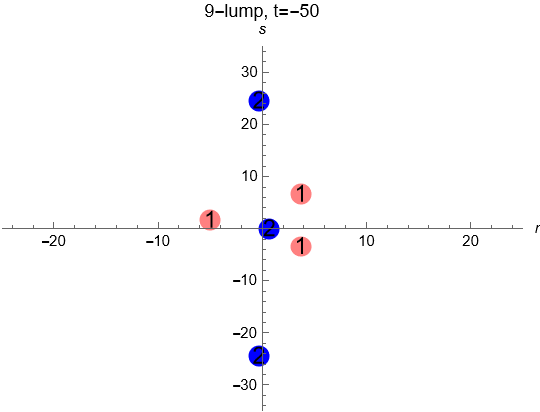}\hfill
\includegraphics[scale=0.45]{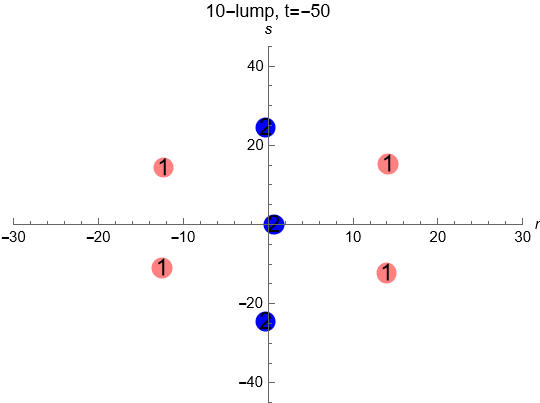}\hfill
\vspace{-0.1cm}{\footnotesize\hspace{1.2cm}(a)\hspace{5.5cm}(b)\hspace{5.2cm}(c)}\\\vspace{0.3cm}
\includegraphics[scale=0.45]{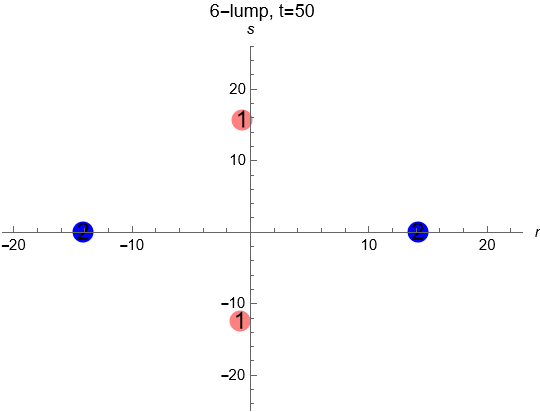}\hfill
\includegraphics[scale=0.45]{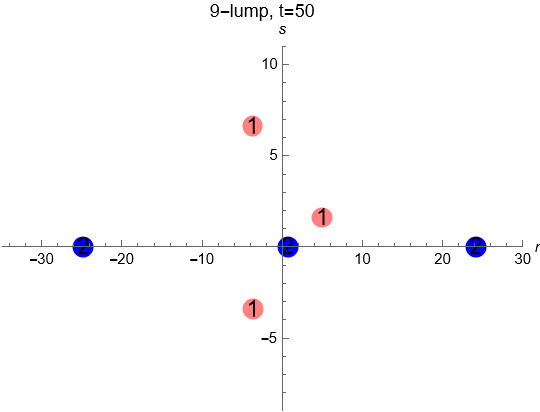}\hfill
\includegraphics[scale=0.45]{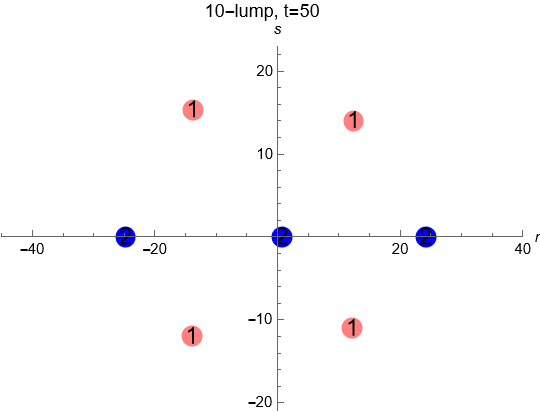}\hfill
\vspace{-0.1cm}{\footnotesize\hspace{1.2cm}(d)\hspace{5.5cm}(e)\hspace{5.2cm}(f)}\\\vspace{0.3cm}
\flushleft{\footnotesize
\textbf{Fig.~$3$.} The parameters are selected as $b=0.5, \omega=0.5$. (a) and (d) show the approximate peak locations of $6$-lump solution. (b) and (e) display the approximate peak locations of $9$-lump solution. (c) and (f) exhibit the approximate peak locations of $10$-lump solution.}
\end{center}

In the figures above, the blue dots represent the multi-peak groups, and the pink dots represent the single-peak groups. Each pink dot represents only the approximate peak position of one peak, while a blue dot represents multiple peaks and the number of peaks is known by the number indicated. Further, we can see that the structures presented in Fig.~$3$ are basically the same as the peak positions in the corresponding examples in Figs.~$1$ and~$2$. In particular, the single-peak groups in the three solutions shown in Fig.~$3$ form a straight line (it can also be viewed as a rectangle), a triangle, and a rectangle, respectively. In the following, we will discuss the approximate distribution of the peak positions by combining the Young diagrams of the different partitions with the corresponding peak positions schematics.

\vspace{2mm}
\textbf{Case A: Rectangular partition}
\vspace{1mm}

Consider the rectangular partition $\lambda=\left(m,m,\cdots,m\right)=\left(m^n\right)$ with $\left|\lambda\right|=N=nm$ and degree vector m=$\left(m,m+1,\cdots,m+n-1\right)$.
The corresponding Young diagram is $m \times n$ rectangle.
Taking $\lambda=\left(4,4,4\right)=\left(4^3\right)$ and $\lambda=\left(4,4,4,4\right)= \left(4^4\right)$ as examples, we can see that when $t>0$, the peaks of all single-peak groups form a rectangle with the same specification as the corresponding Young diagram, while the multi-peak groups are located on the $r$-axis. For $t<0$, it can be viewed approximately as rotated $90^{\circ}$.

\begin{center}
\includegraphics[scale=0.8]{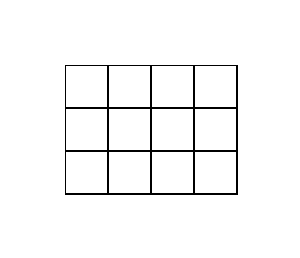}\hfill
\includegraphics[scale=0.5]{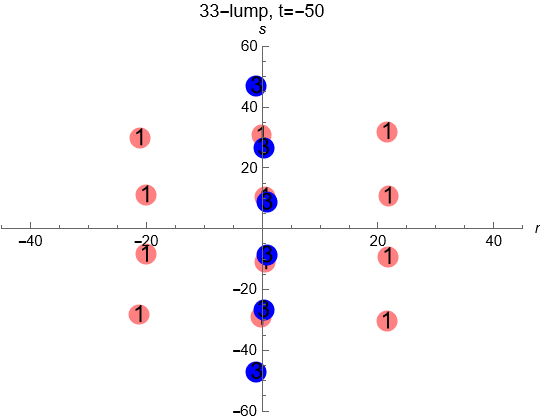}\hfill
\includegraphics[scale=0.5]{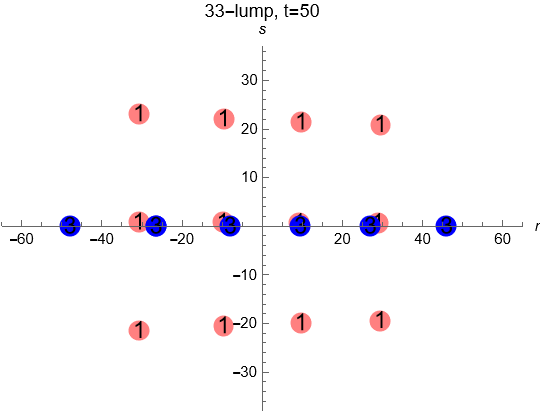}\hfill
\vspace{0cm}{\footnotesize\hspace{2.5cm}(a)\hspace{4.8cm}(b)\hspace{5.3cm}(c)}\\\vspace{0.1cm}
\includegraphics[scale=0.8]{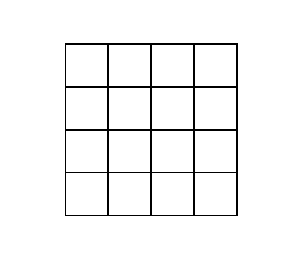}\hfill
\includegraphics[scale=0.5]{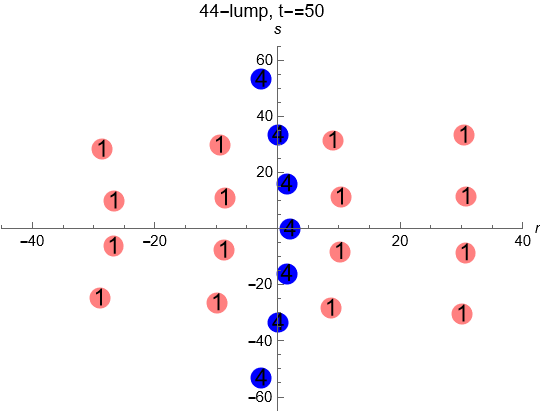}\hfill
\includegraphics[scale=0.5]{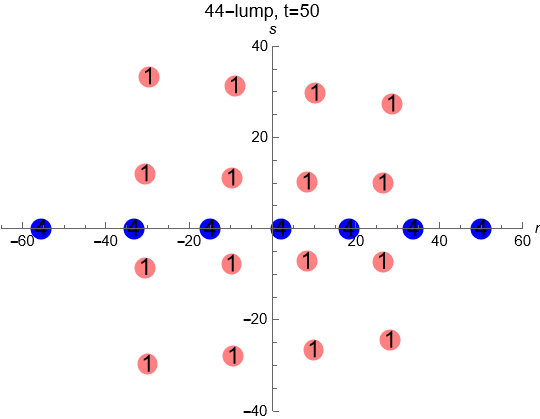}\hfill
\vspace{0cm}{\footnotesize\hspace{2.5cm}(d)\hspace{4.8cm}(e)\hspace{5.3cm}(f)}\\\vspace{0.1cm}
\flushleft{\footnotesize
\textbf{Fig.~$4$.} (a) is the Young diagram of $\lambda=\left(4^3\right)$, (b) and (c) are the peak position graphs for odd partition $\lambda=\left(4^4\right)$ with $b=0.5$ and $\omega=0.5$. (d) is the Young diagram of $\lambda=\left(4^3\right)$, (e) and (f) are the peak position graphs for odd partition $\lambda=\left(4^4\right)$ with $b=0.5$ and $\omega=0.5$.}
\end{center}

\vspace{2mm}
\textbf{Case B: Triangular partition}
\vspace{1mm}

The triangular partition $\lambda=\left(1,2,\cdots,n\right)$ is a conjugate partition with $\left|\lambda\right|=N$, and its corresponding Young diagram is distributed in a staircase shape, also known as a staircase partition. As follows, we take $\lambda=\left(1,2,3\right)$ and $\lambda=\left(1,2,3,4\right)$ as examples to observe the distribution of single-peak groups. When $t<0$, a triangle formed by $N$ single-peak groups points to the negative direction of the $r$-axis, and the opposite is true when $t>0$.

\begin{center}
\includegraphics[scale=0.8]{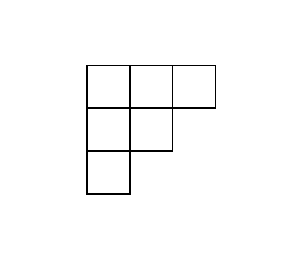}\hfill
\includegraphics[scale=0.5]{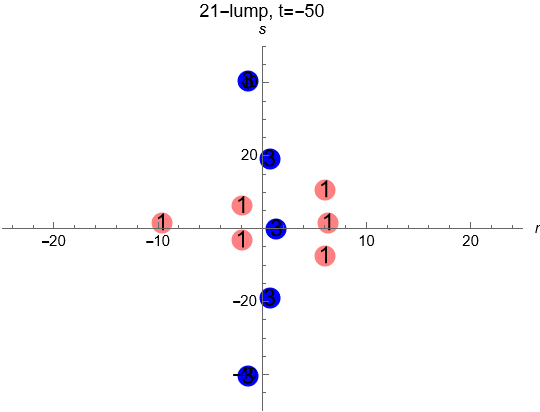}\hfill
\includegraphics[scale=0.5]{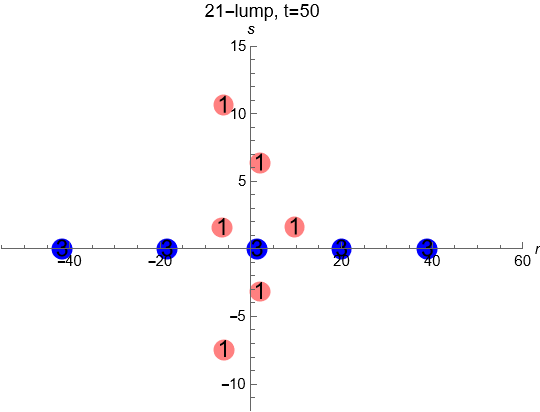}\hfill
\vspace{0cm}{\footnotesize\hspace{2.5cm}(a)\hspace{4.8cm}(b)\hspace{5.3cm}(c)}\\\vspace{0.1cm}
\includegraphics[scale=0.8]{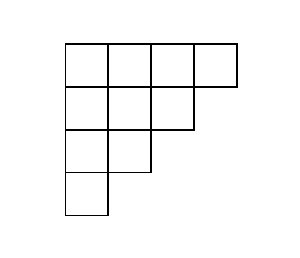}\hfill
\includegraphics[scale=0.5]{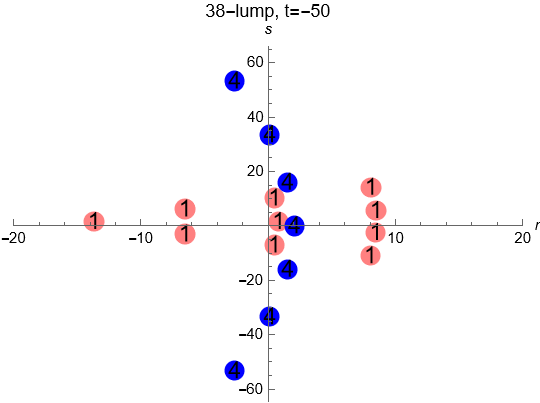}\hfill
\includegraphics[scale=0.5]{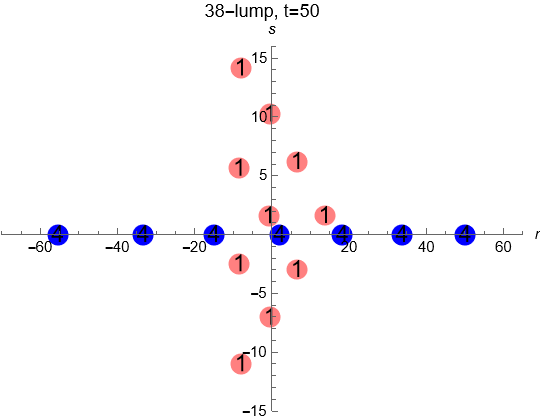}\hfill
\vspace{0cm}{\footnotesize\hspace{2.5cm}(d)\hspace{4.8cm}(e)\hspace{5.3cm}(f)}\\\vspace{0.1cm}
\flushleft{\footnotesize
\textbf{Fig.~$5$.} (a) is the Young diagram of $\lambda=\left(1,2,3\right)$, (b) and (c) are the peak position graphs for odd partition $\lambda=\left(1,2,3,4\right)$ with $b=0.5$ and $\omega=0.5$. (d) is the Young diagram of $\lambda=\left(1,2,3\right)$, (e) and (f) are the peak position graphs for odd partition $\lambda=\left(1,2,3,4\right)$ with $b=0.5$ and $\omega=0.5$.}
\end{center}

\vspace{2mm}
\textbf{Case C: Trapezoidal partition}
\vspace{1mm}

If the partition has the form $\lambda=\left(m,m+1,\cdots,m+n-1\right)$ and $m>1$, then it is called a trapezoidal partition with $\left|\lambda\right|=\frac{n\left(2m+n-1\right)}{2}$.
In particular, when $m=n$ and $\left|\lambda\right|=\frac{n\left(3n-1\right)}{2}$, $\lambda$ is also called a pentagonal partition. In the following, we split the Young diagram of the partition, by removing $d$ copies of two consecutive boxes in each row, so that the new partition $\mu=\left(m-2d,m+1-2d,\cdots,m+n-1-2d\right)$ is a triangle partition.

\begin{center}
\includegraphics[scale=0.8]{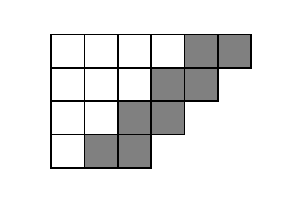}\hfill
\includegraphics[scale=0.5]{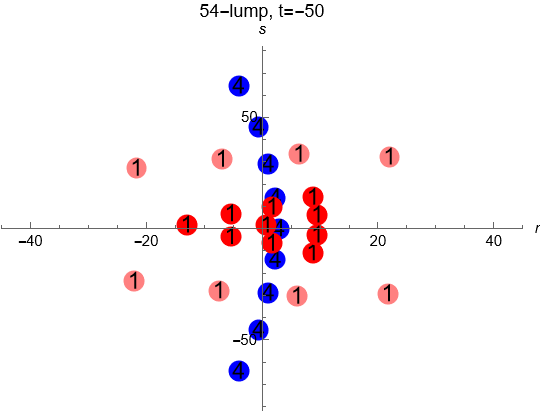}\hfill
\includegraphics[scale=0.5]{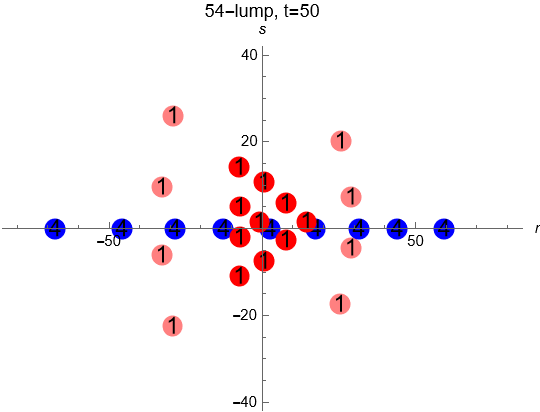}\hfill
\vspace{0cm}{\footnotesize\hspace{2.5cm}(a)\hspace{4.8cm}(b)\hspace{5.3cm}(c)}\\\vspace{0.1cm}
\includegraphics[scale=0.8]{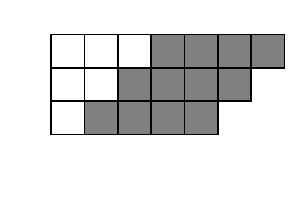}\hfill
\includegraphics[scale=0.5]{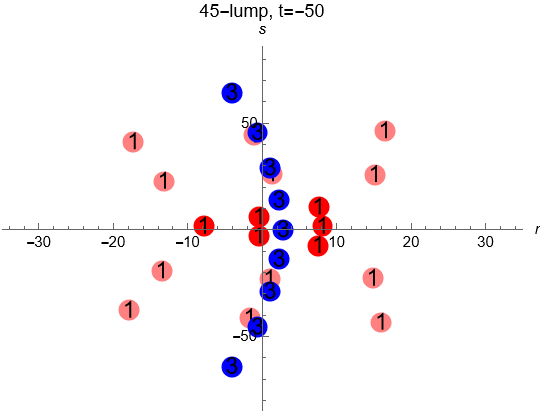}\hfill
\includegraphics[scale=0.5]{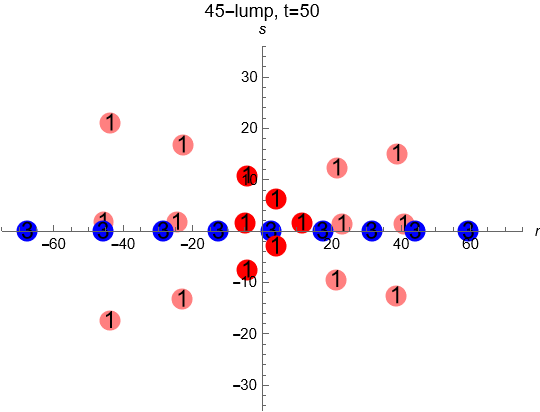}\hfill
\vspace{0cm}{\footnotesize\hspace{2.5cm}(d)\hspace{4.8cm}(e)\hspace{5.3cm}(f)}\\\vspace{0.1cm}
\flushleft{\footnotesize
\textbf{Fig.~$6$.} (a) is the Young diagram of $\lambda=\left(3,4,5,6\right)$, (b) and (c) are the peak position graphs for odd partition $\lambda=\left(5,6,7\right)$ with $b=0.5$ and $\omega=0.5$. (d) is the Young diagram of $\lambda=\left(3,4,5,6\right)$, (e) and (f) are the peak position graphs for odd partition $\lambda=\left(5,6,7\right)$ with $b=0.5$ and $\omega=0.5$.}
\end{center}

\begin{center}
\includegraphics[scale=0.8]{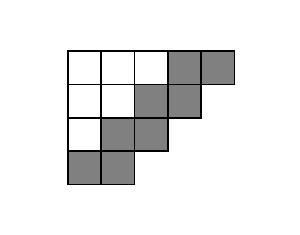}\hfill
\includegraphics[scale=0.5]{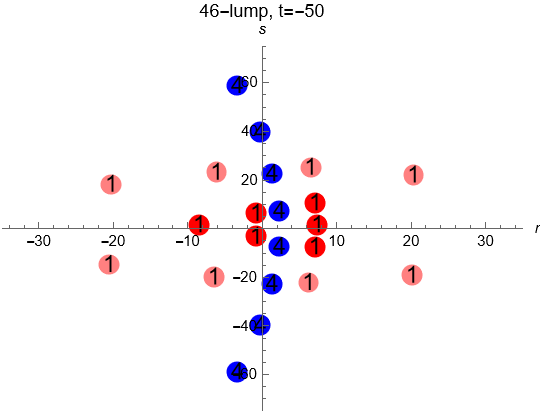}\hfill
\includegraphics[scale=0.5]{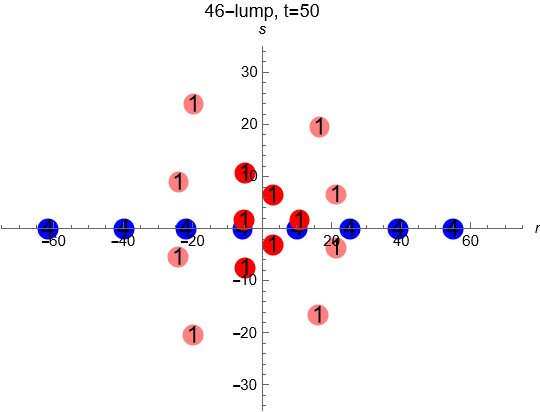}\hfill
\vspace{0cm}{\footnotesize\hspace{2.5cm}(a)\hspace{4.8cm}(b)\hspace{5.3cm}(c)}\\\vspace{0.1cm}
\includegraphics[scale=0.8]{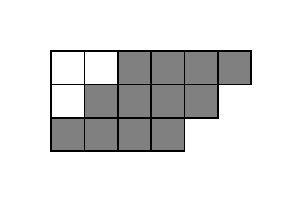}\hfill
\includegraphics[scale=0.5]{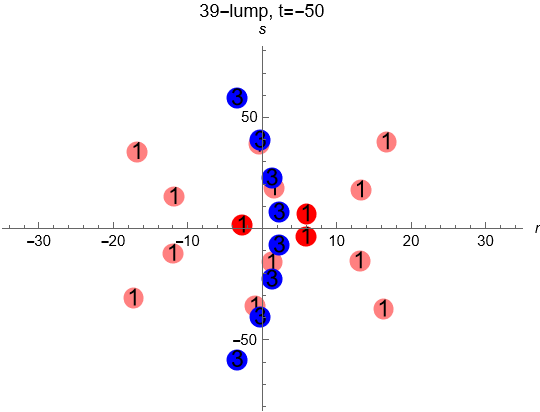}\hfill
\includegraphics[scale=0.5]{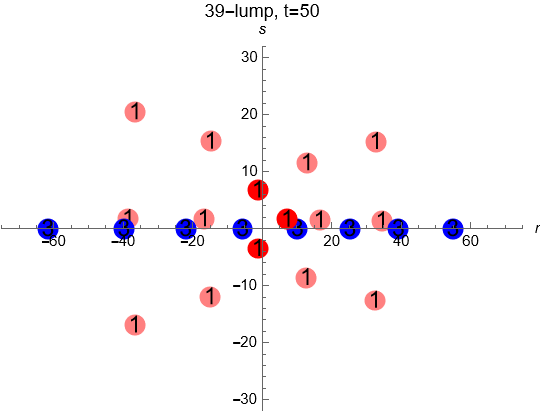}\hfill
\vspace{0cm}{\footnotesize\hspace{2.5cm}(d)\hspace{4.8cm}(e)\hspace{5.3cm}(f)}\\\vspace{0.1cm}
\flushleft{\footnotesize
\textbf{Fig.~$7$.} (a) is the Young diagram of $\lambda=\left(2,3,4,5\right)$, (b) and (c) are the peak position graphs for odd partition $\lambda=\left(4,5,6\right)$ with $b=0.5$ and $\omega=0.5$. (d) is the Young diagram of $\lambda=\left(2,3,4,5\right)$, (e) and (f) are the peak position graphs for odd partition $\lambda=\left(4,5,6\right)$ with $b=0.5$ and $\omega=0.5$.}
\end{center}

When $m$ is odd, the new partition after splitting is $\mu=\left(1,2,\cdots,n\right)$, there are $\left|\mu\right|=\frac{n\left(n+1\right)}{2}$ single-peak groups near the origin forming a triangle pattern, and the remaining $2nd$ single-peak groups are arranged in the periphery of the triangle with symmetric distribution about coordinates, and each side has $d$ arcs composed of $n$ groups. For the case where $m$ is even, the new partition is $\mu=\left(1,2,\cdots,n-1\right)$, similar to the case where $m$ is odd, except that the triangular pattern at the center consists of $\frac{n\left(n-1\right)}{2}$ single-peak groups. Figs.~$6$ and~$7$ correspond to cases where $m$ is odd and even, respectively, and confirm the above conclusion.

\vspace{2mm}
\textbf{Case D: Odd partition}
\vspace{1mm}

If each row of a partition's Young diagram contains an odd number of boxes, then the partition $\lambda=\left(1,3,5,\cdots,2n-1\right)$ is called an odd partition, which has a degree vector of $m=\left(1,4,7,\cdots,3n-2\right)$ and $N=n^2$. For any odd partitions $\lambda=\left(1,3,5,\cdots,2n-1\right)$, we split $q$ boxes in each row of the corresponding Young diagram to form a new diagram, where $q=\frac{\lambda_i-1}{2}, i=1,2,\cdots,n$, and the two new diagrams correspond to two different triangular partitions, which can also be regarded as reorganized into a rectangular partition $\mu=\left(n^n\right)$.
As shown in Fig.~$8$(b), when $t>0$, $N$ single-peak groups form two triangles of side length $n$ with one common side on the $s$-axis. Similarly, it can be also seen as a square with side length $n$, that is, the figure formed by the single-peak groups in Fig.~$4$(e)-(f) is rotated by $45^\circ$ about the central origin.

\begin{center}
\includegraphics[scale=0.8]{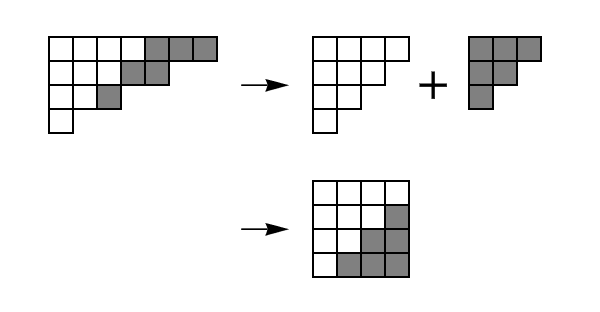}\hfill
\includegraphics[scale=0.5]{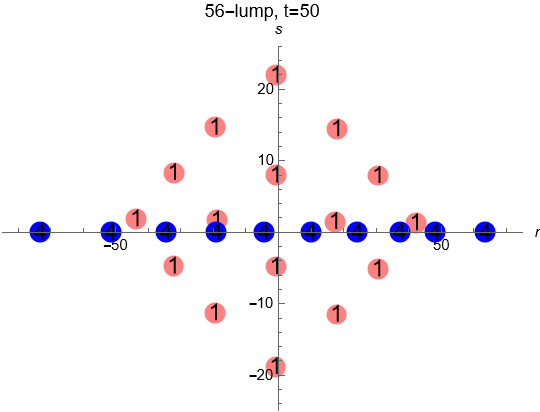}\hfill
\vspace{0cm}{\footnotesize\hspace{4.5cm}(a)\hspace{9.5cm}(b)}\\\vspace{0.1cm}
\flushleft{\footnotesize
\textbf{Fig.~$8$.} (a) is an example of the decomposition of the odd partition $\lambda=\left(1,3,5,7\right)$. (b) is the peak position diagram for odd partition $\lambda=\left(1,3,5,7\right)$ with $b=0.5$ and $\omega=0.5$.}
\end{center}

\vspace{2mm}
\textbf{Case E: Even partition}
\vspace{1mm}

Compared with the Case $D$, even partition $\lambda=\left(2,4,6,\cdots,2n\right)$ corresponds to an even number of boxes in each row. Similarly, after dividing $q=\frac{\lambda_i}{2}$ boxes in each row to form a new graph, the two new Young diagrams correspond to two identical triangular partitions, which can also be regarded as reorganized into a rectangular partition $\mu=\big(\left(n+1\right)^n\big)$.

At $t>0$, the $N$ single-peak groups form two triangles with side length $n$ and whose distribution is symmetric about the $s$ axis. Further, combined with Fig.~$4$(c) and~$9$(b), the shape formed by the single-peak groups corresponding to the even partition can be regarded as the figure formed by the single-peak groups in the converted rectangular partition rotated $45^\circ$ counterclockwise around the origin and split by the $s$-axis.

\begin{center}
\includegraphics[scale=0.8]{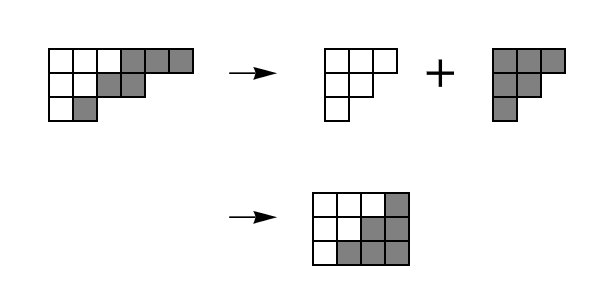}\hfill
\includegraphics[scale=0.5]{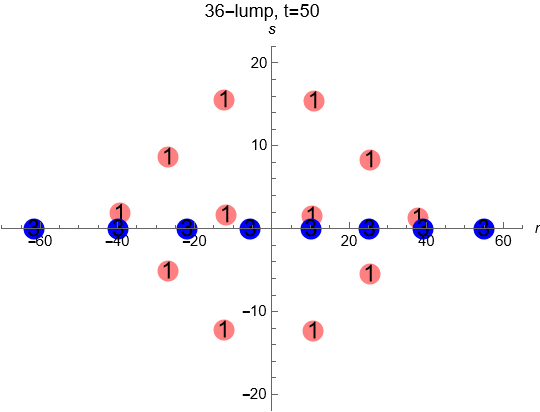}\hfill
\vspace{0cm}{\footnotesize\hspace{4.5cm}(a)\hspace{9.5cm}(b)}\\\vspace{0.1cm}
\flushleft{\footnotesize
\textbf{Fig.~$9$.} (a) is an example of the decomposition of the odd partition $\lambda=\left(2,4,6\right)$. (b) is the peak position diagram for odd partition $\lambda=\left(2,4,6\right)$ with $b=0.5$ and $\omega=0.5$.}
\end{center}

\vspace{2mm}
\textbf{Case F: Square partition}
\vspace{1mm}

If a partition can be denoted as $\lambda=\left(\lambda_1^2,\lambda_2^2, \cdots,\lambda_n^2\right)$, we call it a square partition. This partition can be decomposed into two identical partitions $\lambda=\left(\lambda_1,\lambda_2, \cdots,\lambda_n\right)$, and the graph formed by the single-peak groups is symmetric with respect to the $r$ axis when $t>0$, as verified by Fig.~$10$(b) and~(d). For partition $\lambda=\left(3^2,4^2\right)$, we can obtain two triangles with side lengths of $2$, symmetrically positioned on the $r$-axis, and creating two arcs around the triangle, as depicted in the bottom half of Fig.~$10$(c).

\begin{center}
\includegraphics[scale=1]{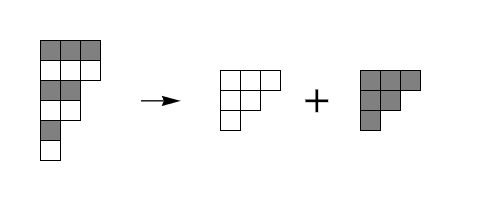}\hfill
\includegraphics[scale=0.5]{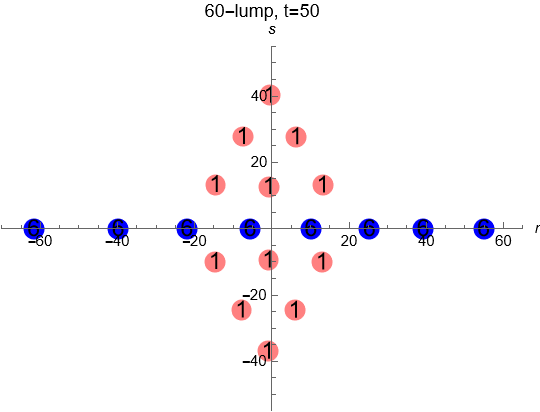}\hfill
\vspace{0cm}{\footnotesize\hspace{4.5cm}(a)\hspace{9.5cm}(b)}\\\vspace{0.1cm}
\includegraphics[scale=0.8]{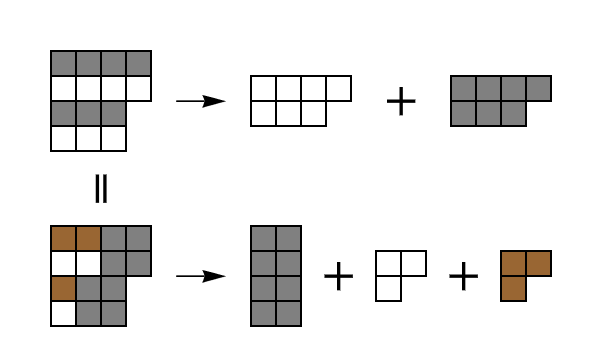}\hfill
\includegraphics[scale=0.5]{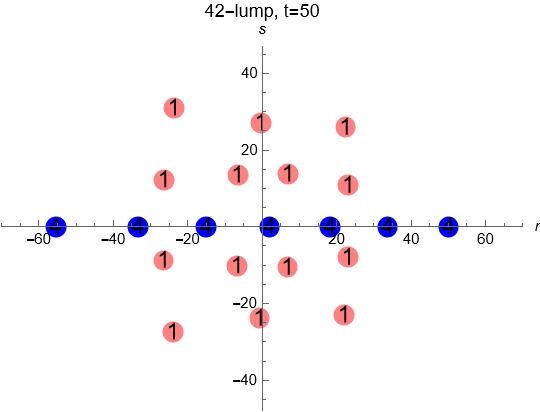}\hfill
\vspace{0cm}{\footnotesize\hspace{4.5cm}(c)\hspace{9.5cm}(d)}\\\vspace{0.1cm}
\flushleft{\footnotesize
\textbf{Fig.~$10$.} (a) is an example of the decomposition of the odd partition $\lambda=\left(1^2,2^2,3^2\right)$. (b) is the peak position diagram for odd partition $\lambda=\left(1^2,2^2,3^2\right)$ with $b=0.5$ and $\omega=0.5$. (c) is an example of the decomposition of the odd partition $\lambda=\left(3^2,4^2\right)$. (d) is the peak position diagram for odd partition $\lambda=\left(3^2,4^2\right)$ with $b=0.5$ and $\omega=0.5$.}
\end{center}

Combined with the above special partitions and corresponding examples, we can draw the following conclusions.

\textbf{(i)} For any partition $\lambda=\left(\lambda_1,\lambda_2, \cdots,\lambda_n\right)$, whose degree vector is $m=\left(m_1,m_2, \cdots,m_n\right)$, the multi-lump solution generated by the partition contains $M=N+nm_n$ peaks, which can be divided into $m_n$ multi-peak groups and $N=\left|\lambda\right|$ single-peak groups, and each multi-peak group contains $l\left(\lambda\right)=n$ peaks. In terms of position distribution, when $t>0$, the multi-peak groups are located on the $r$-axis, and when $t<0$, it is approximately parallel to the $s$-axis.

\textbf{(ii)} Take the splitting process of $t>0$ as an example, if the Young diagram of partition is split into left and right parts, such as odd partition and even partition, the graph formed by single-peak groups of the partition is also about the $s$-axis distribution. If it is split by row, the graph formed by the single-peak groups is distributed up and down the $r$-axis, such as square partition.

\textbf{(iii)} The graph formed by the single-peak groups of partition $\lambda$ at $t>0$ is the same as that of the conjugate partition $\lambda^{\prime}$ at $t<0$.

\textbf{(iv)} For a class of partitions with only one triangle in the graph formed by single-peak groups, such as triangular partitions and trapezoidal partitions, the triangle points in the positive direction of the $r$-axis when $t>0$, and the opposite direction when $t<0$. For odd and even partitions, the single-peak groups form two triangles pointing in both directions of the $r$-axis when $t>0$, and in the positive and negative directions of the $s$-axis when $t<0$.

Through the discussion in this section, we obtained the approximate locations and distribution characteristics of the peaks of multi-lump solutions formed by different integer partitions, especially a special class of partitions, which will provide a solid foundation for us to further understand the characteristics of DLWEs and the application in different fields.

\vspace{5mm}\noindent\textbf{5  Conclusions}\\
\hspace*{\parindent}

In this paper, we mainly studied the relationship between the multi-lump solutions of DLWEs and integer partition, and the following conclusions were obtained. By introducing generalized Schur polynomials and the integer partition theory, combined with rational solutions constructed by the BDT, the multi-lump solutions of DLWEs have been derived. Furthermore, we have studied the distribution of peak positions of lump solutions with special partition, especially the distribution characteristics of single-peak groups, and have summarized the relevant properties of the distribution of peak positions. In general, this paper further analyzed the peak properties of multi-lump solutions obtained under the condition of high-order BDT, which not only expanded our understanding of DLWEs, but also provided a theoretical basis for other fields of research and application. However, there is still much unexplored research on the multi-lump solutions of DLWEs, which is a direction we are still working on.

\vspace{5mm}\noindent\textbf{Acknowledgments}\\
\hspace*{\parindent}

We express our sincere thanks to each member of our discussion group for their suggestions. This work has been supported by the Shanxi Province Science Foundation under Grant No. 202303021221031, and the Fund Program for the Scientific Activities of Selected Returned Overseas Scholars in Shanxi Province under Grant No. 20220008.

\vspace{5mm}\noindent\textbf{Data availability}\\
\hspace*{\parindent}

Enquiries about data availability should be directed to the authors.

\vspace{5mm}\noindent\textbf{Conflict of interest}\\
\hspace*{\parindent}

The authors have not disclosed any competing interests.


\begin{thebibliography}{99}


\bibitem{ck81}
R. Grimshaw, E. Pelinovsky, T. Taipova, A. Sergeeva. Rogue internal waves in the ocean: Long wave model. {\it Eur. Phys. J. Spec. Top.} {\bf 185(1)} (2010) 195-208.

\bibitem{ck82}
Z. I. Fedotova, G. S. Khakimzyanov. Nonlinear-dispersive shallow water equations on a rotating sphere. {\it Russ. J. Numer. Anal. Math. Model.} {\bf 25(1)} (2010) 15-26.

\bibitem{ck83}
X. Y. Gao, Y. J. Guo, W. R. Shan. Oceanic shallow-water symbolic computation on a $\left(2 + 1\right)$-dimensional generalized dispersive long-wave system. {\it Phys. Lett. A} {\bf 457} (2023) 128552.

\bibitem{ck84}
M. Niwas, S. Kumar. New plenteous soliton solutions and other form solutions for a generalized dispersive long-wave system employing two methodological approaches. {\it Opt. Quantum Electron.} {\bf 55(7)} (2023) 630.

\bibitem{ck85}
M. S. Aktar, M. A. Akbar, K. S. Nisar, H. I. Alrebdi, A. Abdel-Aty. Steeping and dispersive effects analysis of a couple of long-wave equations in dispersive media. {\it Alex. Eng. J.} {\bf 61(12)} (2022) 9457-9470.

\bibitem{ck86}
T. B. Benjamin, J. L. Bona, J. J. Mahony. Model equations for long waves in nonlinear dispersive systems. {\it Phil. Trans. R. Soc. A} {\bf 272(1220)} (1972) 47-78.

\bibitem{ck87}
C. Q. Dai, Y. Y. Wang, A. Biswas. Dynamics of dispersive long waves in fluids. {\it Ocean Eng.} {\bf 81} (2014) 77-88.

\bibitem{ck1}
M. Boiti, J. J. P. Leon, F. Pempinelli. Spectral transform for a two spatial dimension extension of the dispersive long wave equation. {\it Inverse Probl.} {\bf 3(3)} (1987) 371-387.

\bibitem{ck31}
C. Q. Dai, F. D. Zong, J. F. Zhang. Nonpropagating solitary waves in $\left(2 + 1\right)$-dimensional generalized dispersive long wave systems. {\it Int. J. Theor. Phys.} {\bf 45} (2006) 790-801.

\bibitem{ck32}
Y. H. Tian, H. L. Chen, X. Q. Liu. New exact solutions to dispersive long-wave equations in $\left(2 + 1\right)$-dimensional space. {\it Commun. Theor. Phys.} {\bf 45(2)} (2006) 207-210.

\bibitem{ck33}
L. L. Feng, S. F. Tian, T. T. Zhang. Nonlocal symmetries and consistent Riccati expansions of the $\left(2 + 1\right)$-dimensional dispersive long wave equation. {\it Z. Naturforsch. A} {\bf 72(5)} (2017) 425-431.

\bibitem{ck34}
J. Hu, Z. W. Xu, G. F. Yu. Determinant structure for the $\left(2 + 1\right)$-dimensional dispersive long wave system. {\it Appl. Math. Lett.} {\bf 62} (2016) 76-83.

\bibitem{ck35}
H. Y. Zhang, Y. F. Zhang. Rational solutions and their interaction solutions for the $\left(2 + 1\right)$-dimensional dispersive long wave equation. {\it Phys. Scr.} {\bf 95(4)} (2020) 045208.

\bibitem{ck36}
Z. Y. Yan. The investigation for $\left(2 + 1\right)$-dimensional Eckhaus-type extension of the dispersive long wave equation. {\it J. Phys. A: Math. Gen.} {\bf 37(2)} (2004) 841.

\bibitem{ck37}
H. Wang, Y. H. Wang, H. H. Dong. Interaction solutions of a $\left(2 + 1\right)$-dimensional dispersive long wave system. {\it Comput. Math. Appl.} {\bf 75(8)} (2018) 2625-2628.

\bibitem{ck38}
M. Eslami. Solutions for space-time fractional $\left(2 + 1\right)$-dimensional dispersive long wave equations. {\it Iran. J. Sci. Technol. A} {\bf 41} (2017) 1027-1032.

\bibitem{ck39}
Y. Q. Zhou, Q. Liu. Bifurcation of travelling wave solutions for a $\left(2 + 1\right)$-dimensional nonlinear dispersive long wave equation. {\it Appl. Math. Comput.} {\bf 189(1)} (2007) 970-979.

\bibitem{ck40}
Z. Y. Yan. Generalized transformations and abundant new families of exact solutions for $\left(2 + 1\right)$-dimensional dispersive long wave equations. {\it Comput. Math. Appl.} {\bf 46(8-9)} (2003) 1363-1372.


\bibitem{ck20}
Y. L. Dang, H. J. Li, J. Lin. Soliton solutions in nonlocal nonlinear coupler. {\it Nonlinear Dyn.} {\bf 88} (2017) 489-501.

\bibitem{ck21}
S. Singh, K. Sakkaravarthi, T. Tamizhmani, K. Murugesan. Painlev\'{e} analysis and higher-order rogue waves of a generalized $\left( 3+1 \right)$-dimensional shallow water wave equation. {\it Phys. Scr.} {\bf 97(5)} (2022) 055204.

\bibitem{ck22}
Z. L. Zhao, L. C. He, A. M. Wazwaz. Dynamics of lump chains for the BKP equation describing propagation of nonlinear waves. {\it Chin. Phys. B} {\bf 32(4)} (2023) 040501.

\bibitem{ck23}
T. H. Andriotty, P. S. Schneider, L. J. Rodrigues. Accuracy of lumped element model for cyclic sensible thermal energy storage systems. {\it J. Energy Storage} {\bf 28} (2020) 101277.

\bibitem{ck24}
Y. F. Jian, F. W. Bai, Q. Falcoz, C. Xu, Y. Wang, Z. F. Wang. Thermal analysis and design of solid energy storage systems using a modified lumped capacitance method. {\it Appl. Therm. Eng.} {\bf 75} (2015) 213-223.

\bibitem{ck25}
H. Leblond, M. Manna. Nonlinear dynamics of two-dimensional electromagnetic solitons in a ferromagnetic slab. {\it Phys. Rev. B} {\bf 77(22)} (2008) 224416.

\bibitem{ck26}
K. Q. Li. Nonlinear dynamics for different nonautonomous wave structure solutions. {\it Open Phys.} {\bf 20(1)} (2022) 464-469.



\bibitem{ck10}
H. Q. Zhao, W. X. Ma. Mixed lump-kink solutions to the KP equation. {\it Comput. Math. Appl.} {\bf 74(6)} (2017) 1399-1405.

\bibitem{ck11}
C. Lester, A. Gelash, D. Zakharov, V. Zakharov. Lump chains in the KP-I equation. {\it Stud. Appl. Math.} {\bf 147(4)} (2021) 1425-1442.

\bibitem{ck12}
S. Chakravarty, M. Zowada. Multi-lump wave patterns of KPI via integer partitions. {\it Physica D} {\bf 446} (2023) 133644.

\bibitem{ck13}
S. Chakravarty. Multi-lump solutions of KPI. {\it Nonlinear Dyn.} {\bf 112(1)} (2024) 575-589.

\bibitem{ck62}
V. B. Matveev. Some comments on the rational solutions of the Zakharov-Schabat equations. {\it Lett. Math. Phys.} {\bf 3} (1979) 503-512.

\bibitem{ck63}
D. Pelinovsky. Rational solutions of the Kadomtsev-Petviashvili hierarchy and the dynamics of their poles. I. New form of a general rational solution. {\it J. Math. Phys.} {\bf 35(11)} (1994) 5820-5830.

\bibitem{ck61}
D. Pelinovsky. Rational solutions of the KP hierarchy and the dynamics of their poles. II. Construction of the degenerate polynomial solutions. {\it J. Math. Phys.} {\bf 39(10)} (1998) 5377-5395.

\bibitem{ck65}
S. Chakravarty, M. Zowada. Dynamics of KPI lumps. {\it J. Phys. A: Math. Theor.} {\bf 55(19)} (2022) 195701.

\bibitem{ck66}
P. G. L. Leach. Heat polynomials and Lie point symmetries. {\it J. Phys. A: Math. Theor.} {\bf 322(1)} (2006) 288-297.

\bibitem{ck64}
P. C. Rosenbloom, D. V. Widder. Expansions in terms of heat polynomials and associated functions. {\it Trans. Am. Math. Soc.} {\bf 92(2)} (1959) 220-266.


\end{thebibliography}
\end{document}